\documentclass[10pt]{article}

\usepackage{amsmath}  
\usepackage{graphicx}
\usepackage{caption}
\usepackage{color}
\usepackage[margin=0.5in]{geometry}
\usepackage{multirow}
\usepackage{tabularx}
\usepackage{cite}
\usepackage[font=footnotesize,labelfont=bf]{caption}
\usepackage{float}
\usepackage{amssymb}
\usepackage{amsmath}
\usepackage{subfig}
\usepackage{multicol}

\title{\bf{Scattering in terms of Bohmian conditional wave functions for scenarios with non-commuting energy and momentum operators}}
\author{{Matteo Villani}$^{1}$, {Guillermo Albareda}$ ^{2,3}$, {Carlos Destefani}$^{1}$,  {Xavier Cartoixà}$^{1}$, {Xavier Oriols}$^{1}$}
\date{\small{1: Department of Electronic Engineering, Universitat Aut\`onoma de Barcelona, Campus de la UAB, 08193 Bellaterra, Barcelona, Spain\\ 2: Max Planck Institute for the Structure and Dynamics of Matter, Luruper Chaussee 149, 22761 Hamburg, Germany \\ 3: Institute of Theoretical and Computational Chemistry, Universitat de Barcelona, Gran Via de les Corts Catalanes 585, 08007 Barcelona, Spain}}

\begin{document}
\maketitle
\begin{abstract}
Without access to the full quantum state, modeling quantum transport in mesoscopic systems requires dealing with a limited number of degrees of freedom. In this work, we analyze the possibility of modeling the perturbation induced by the non-simulated degrees of freedom on the simulated ones as a transition between single-particle pure states. First, we show that Bohmian conditional wave functions (BCWF) allow a rigorous discussion of the dynamics of electrons inside open quantum systems in terms of such single-particle pure states, either under Markovian or non-Markovian conditions. Second, we discuss the practical application of the method for modeling light-matter interaction phenomena in a resonant tunneling device (RTD), where a single photon is interacting with a single electron. Third, we emphasize the importance of interpreting such scattering mechanism as a transition between initial and final single-particle BCWF with well-defined energies (rather than with well-defined momenta).
\end{abstract}

\paragraph{keywords:}$\!\!$quantum dissipation; Bohmian mechanics; collision; conditional wave function; decoherence; open systems; many-body problem
\paragraph{Corresponding author:}$\!\!$matteo.villani@uab.es

\begin{multicols}{2}
\section{Introduction}
\label{intro}

Due to the well-known many-body problem, electron transport in nanoscale devices has to be modeled as an open quantum system~\cite{Start}. The contacts, cables, atoms, electromagnetic radiation, etc.\ are commonly considered to be part of the environment. The effect of this environment on the dynamics of the simulated degrees of freedom, i.e., the electrons in the active region, can be recovered using some kind of perturbative approximation. There are different formalisms in the literature to deal with such \emph{environmental} perturbation (Green's functions \cite{Green,Klimeck,Klimeck2}, density matrix \cite{Rossi1,Rossi2}, Wigner distribution function \cite{Wigner,Ferry,Frensley,Dollfus1,Dollfus2}, Kubo formalism \cite{Cummings}, Pauli quantum Master equation \cite{Fischetti1, Fischetti2}, pure states \cite{Kramer1,Kramer2}, etc). In this work we analyze the possibility of modeling the quantum nature of such simulated degrees of freedom with single-particle pure states and the \textit{environmental} perturbations on the simulated degrees of freedom by the non-simulated ones as a transition between initial and final single-particle pure states.

In particular, we are interested in modeling the collision of an electron with a phonon or/and photon in an active region with tunneling barriers, i.e. in a scenario where the energy and momentum operators do not commute. The path for achieving this goal requires first addressing the answer of the following question: \textit{Is it possible to model an open system in terms of single-particle pure states?}. Once this conceptual question is answered, the next practical question that needs to be addressed is: \textit{How do we select the single-particle pure states before and after the collision?} In this paper we answer both questions. It will be shown that the alternative Bohmian formulation of quantum transport  \cite{Bohm_original} provides a rigorous and versatile tool to describe collisions in open quantum systems in terms of single-particle pure wave functions. This work is part of a long-term research project for the development of a general-purpose nanoelectronic device simulator, the so-called BITLLES simulator \cite{BITLLES1}, using Bohmian trajectories.

The structure of the paper is the following. In Sec. \ref{s2} the answer to the first question about using single-particle pure states for open systems is provided from the Bohmian description of quantum phenomena. In Sec. \ref{s3} we provide an exact model for matter-light interaction in a closed system, and some simulation results are reported for different conditions of the total energy of the closed system, with a final discussion on the interaction between active region and environment, to extend this description to an open system. In Sec. 4, the practical implementation of the transition between a pres-selected $|i\rangle$ and a post-selected $|f\rangle$ states: $|i\rangle \to |f\rangle$ is discussed. This done for two different models: model A deals with the energy conservation and model B deals with the momentum conservation. In Sec. \ref{s5} these two models are compared, for a flat potential and for an arbitrary potential, to each other and to the exact model of Sec. \ref{s3}. Our conclusions are summarized in Sec. \ref{s6}.

\section{Is it possible to model open system in terms of single-particle pure states?}
\label{s2}

As we have stated the active region of an electron device is, strictly speaking, an open quantum system interacting with the contacts, atoms in thermal motion, radiation, etc.  As a consequence, in principle one is not allowed to describe the electron in the active region in terms of pure states, but one has to rely on the use of the reduced density matrix.

Most approaches to open systems revolve around the reduced density matrix built by tracing out the degrees of freedom of the environment \cite{Start}. The ability to describe open systems with pure states can be partially justified when dealing with Markovian systems. In a pragmatical definition of Markovianity \cite{Markovianity_overview}, the correlations between system and environment decay in a time scale that is much smaller than the observation (or relevant) time interval of the system. Thus, it can be assumed that every time we observe the system it is defined by a pure state. For Markovian evolutions, the Lindblad master equation \cite{Lindblad} for the reduced density matrix is a standard simulation tool. In addition, in Markovian scenarios where the off-diagonal terms of the reduced density matrix become irrelevant, a quantum master equation can be implemented dealing with transitions between pure states \cite{Fischetti1, Fischetti2}.

In fact, it is possible to develop stochastic Schr\"odinger equations to unravel the reduced density matrix in terms of a pure-state solution for either Markovian or non-Markovian systems. The pure-state solution of stochastic Schr\"odinger equations can be interpreted as the state of the Markovian system while the environment is under (continuous) observation. However, such physical interpretation cannot be given to the solutions of the stochastic Schr\"odinger equations for non-Markovian systems \cite{vega,GRW1,GRW2,SSE1,SSE2,SSE3,Gambetta1,Gambetta2,Diosi,ferialdi}, where pure states can provide the correct one-time ensemble value, but cannot be used to compute time correlations.

Therefore, for general non-Markovian quantum processes, when we are interested in a time-resolved description of the electron device performance, it is not possible to define the open system in terms of orthodox pure states.  As described in \cite{PRLxavier} and explained below, a proper solution for treating electrons in non-Markovian open systems as single-particle pure states comes from the Bohmian formalism.

To explain how the Bohmian theory allows a general description of a many-body quantum system in terms of wave functions, we consider a simplified scenario with only two degrees of freedom: one degree of freedom $x$ belonging to the system, plus one degree of freedom $y$ belonging to the environment. Thus, the pure state in the position representation solution of the unitary Schrodinger equation is $\Psi(x,y,t)$. For each experiment, labelled by $j$, a Bohmian quantum state is defined by this $\Psi(x,y,t)$ plus two well-defined trajectories,  $X^j[t]$ in the $x$-physical space and $Y^j[t]$ in the $y$-physical space. The role of the many-body wavefunction $\Psi(x,y,t)$ is guiding each trajectory $X^j[t]$ with a velocity that reads~\cite{Bohm_original, Bohm1, BookXavier}
\begin{eqnarray}
\label{velo}
v_x^j[t]=\frac{d X^j[t]}{dt}=\frac { J_x(X^j[t],Y^j[t],t)}{|\Psi(X^j[t],Y^j[t],t)|^2} =\frac{1}{m^*}\left.\frac{\partial S(x,y,t)}{\partial x}\right|_{x=X^j[t],y=Y^j[t]},
\end{eqnarray}
where $J_ x(x,y,t) = \hbar \mathop{\rm Im} \left(\Psi^*(x,y,t) \frac{\partial}{\partial_x} \Psi(x,y,t)) \right)/m^*$ is the current density with $m^*$ the mass of the $x$-particle, and $S(x,y,t)$ is the phase of the wave function written in polar form $\Psi(x,y,t)=|\Psi(x,y,t)|e^{iS(x,y,t)/\hbar}$. Analogous definitions are possible for the $Y^j[t]$ trajectory. The two positions $\{X^j[t],Y^j[t]\}$ in different $j=1,...,W$ experiments are distributed (obeying quantum equilibrium \cite{Bohm1, BookXavier}) as
\begin{eqnarray}
\label{QE}
|\Psi(x,y,t)|^2 =  \frac{1}{W} \sum_{j=1}^{W} \delta(x-X^j[t]) \delta(y-Y^j[t]).
\end{eqnarray}
The identity in \eqref{QE} requires $W \to \infty$. Numerically, we will just require a large enough $W$ to reproduce ensemble values given by the Born law in agreement with the orthodox theory.

The Bohmian theory opens the possibility to deal with a wave function of a subsystem, through the concept of Bohmian conditional wave function (BCWF) \cite{BookXavier, Enrique1}. The BCWF is defined for the $x$-degree of freedom during the $j$-th experiment as
\begin{equation}
\psi^j(x,t)\equiv \Psi(x,Y^j[t],t).
\label{cond}
\end{equation}
We emphasize that $\psi^j(x,t)$ provides a rigorous (Bohmian) definition of a single-particle wave function for an open system \cite{Bohm1} that still includes the correlations with the other degree of freedom $y$. Notice that the reason why the BCWF has a relevant role in the Bohmian theory is because the trajectory $X^j[t]$ is equivalently guided by $\Psi(x,y,t)$ or by $\psi^j(x,t)$. In other words, the velocity $v^j_x[t]$ in \eqref{velo} can be equivalently computed from the wave function as
\begin{eqnarray}
\label{velo2}
v^j_x[t]=\frac{d X^j[t]}{dt}=\frac { J_x^j(X^j[t],t)}{|\psi^j(X^j[t],t)|^2}=\frac{1}{m^*}\left.\frac{\partial s(x,t)}{\partial x}\right|_{x=X^j[t]} ,
\end{eqnarray}
where   $|\psi^j(X^j[t],t)|^2=|\Psi(X^j[t],Y^j[t],t)|^2$, $J_ x^j(x,t) = \hbar \mathop{\rm Im} \left(\psi^{j,*}(x,t) \frac{\partial}{\partial_x} \psi^j(x,t)) \right)/m^*$, and $s(x,t)$ is the angle of the BCWF in polar form $\psi^j(x,t)=|\psi^j(x,t)|e^{is(x,t)/\hbar}$. Notice that we have not done any approximation about the Markovianity of the quantum system in the definition of the BCWF. Thus, at the conceptual level, we conclude that any quantum open system can be analyzed in terms of single-particle pure states (i.e., BCWF) using the Bohmian formalism. This is a well-known result \cite{PRLxavier} and provides a definitive positive answer to the initial question: \textit{Is it possible to model open system in terms of single-particle pure states?} 

Let us discuss now a more realistic scenario with $N$ electrons inside the active region with degrees of freedom $\{x_1,x_2,...,x_N\}$ that we want to simulate explicitly (for simplicity each electron is assumed to be defined in a 1D space). There are, however, $M$ environmental degrees of freedom $\{y_1,y_2,...,y_M\}$that we do not want to simulate explicitly. The new many-body wave function of such scenario is $\Psi(x_1,x_2,...,x_N,y_1,y_2,...,y_M)$, which is numerically inaccessible. We define $\bar X^j_i[t]=\{x^j_1[t],..,x_{i-1}[t],x^j_{i+1}[t],...,x^j_N\}$ as the set of all Bohmian trajectories of the system except $x^j_i(t)$ for the $j$-experiment. Notice that we are dealing now with a superindex $j$ indicating the experiment and  subindex $i$ indicating each particle in a given experiment. We also define $Y^j[t]=\{y^j_1[t],.....,y^j_M[t]\}$ as the set of all trajectories of the environment at the $j$-experiment. Then, the set of equations of motion of the resulting $N$ single-electron conditional wave functions $\psi^j(x_1,t) \equiv \Psi(x_1,\bar X^j_1[t],Y^j[t],t),...,\psi^j(x_N,t) \equiv \Psi(x_N,\bar X^j_N[t],Y^j[t],t)$ inside the active region can be written as:
\begin{eqnarray}\label{EQM_conditional5}
i\hbar\frac{d\psi^j(x_1,t)}{dt}& =&  \left[-\frac{\hbar ^2}{2m}\nabla^2_{x_1}+U_{eff}^j(x_1,t)\right] \psi^j(x_1,t)\nonumber\\
&\vdots&    \\
i\hbar\frac{d\psi^j(x_N,t)}{dt} &=&  \left[-\frac{\hbar ^2}{2m}\nabla^2_{x_N}+U_{eff}^j(x_N,t)\right] \psi^j(x_N,t).\nonumber
\end{eqnarray}
The effective single-particle potential $U_{eff}^j(x_i,t)\equiv U_{eff}^j(x_i,\bar X^j_i[t],Y^j[t],t)$ is
\begin{eqnarray}
U_{eff}^j(x_i,t)&=&{U}^j(x_i,t)+ V^j(x_i,t) \nonumber \\ &+&\mathcal{A}^j(x_i,t)+i\mathcal{B}^j(x_i,t),
\label{potentials}
\end{eqnarray}
where $U^j(x_i,t)$ is an external potential acting only on the system degrees of freedom $x_i$, $V^j(x_i,t)$ is the Coulomb potential between $x_i$ and the rest of particles at fixed positions $\bar X^j_1[t]$ and $Y^j[t]$, and $\mathcal{A}^j(x_i,t)$ and $\mathcal{B}^j(x_i,t)$ are potentials responsible for the remaining of quantum correlations between the degrees of freedom of the system and the environment~\cite{PRLxavier}. A mandatory clarification is needed here. Are the set of the BCWF in \eqref{EQM_conditional5} solving the many-body problem? No. If you want to use the coupled system of equations of motion of the $N$ BCWF in \eqref{EQM_conditional5} to describe a given experiment, first, you have to solve the Poisson (Gauss) equation to find $U^j(x_i,t)$ and $V^j(x_i,t)$ explicitly and, second, you have to know the exact solution of the many-body wave function $\Psi(x_1,x_2,...,x_N,y_1,y_2,...,y_M)$ to find $\mathcal{A}^j(x_i,t)$ and $\mathcal{B}^j(x_i,t)$ for all electrons\cite{PRLxavier}. The last step is numerically inaccessible. The merit of the system of equations in \eqref{EQM_conditional5} is showing that such type of solutions of the many-body function exists and that we can look for educated guesses on the shape of $\mathcal{A}^j(x,t)$ and $\mathcal{B}^j(x,t)$ to provide reasonable approximations. Notice that a similar procedure is followed in Density Functional Theory: it shows a way of rewriting the many-body wave function in terms of single-particle wave functions, but the procedure requires the knowledge of the exchange-correlation functional, which is only known once the many-body wave function is known. See further details and an explanation on $\mathcal{A}^j(x,t)$ and $\mathcal{B}^j(x,t)$ in \cite{BookXavier, PRLxavier,Bohm1,BITLLES1, Albareda, Marian}.

To better appreciate the details of this simulation technique for electron devices, we notice that the total current $I^j(t)$ at time $t$ for the $j$-experiment, after solving the set of BCWF from \eqref{EQM_conditional5} with the appropriate approximations for $\mathcal{A}^j(x,t)$ and $\mathcal{B}^j(x,t)$ can be defined from the Bohmian trajectories with the help of a quantum version of the Ramo--Schokley--Pellegrini theorem~\cite{albareda2012computation} as:
\begin{equation}\label{ramo}
I^j(t)=  \frac{e}{L} \sum_{i=1}^{n(t)} v_{x_i}^j(x_i^j[t],\bar X^j_i[t],Y^j[t]),
\end{equation}
where $L$ is the distance between the two (metallic) contacts, $e$ is the electron charge, and $v_{x_i}(x^j_i[t],\bar X^j_i[t],Y^j[t] t)$ is the Bohmian velocity of the $i$-th electron inside the active region in the $j$-experiment.  Notice that the \textit{observables} are computed from the trajectories (not from the BCWF) and they are linked to a particular experiment $j$ (which can be understood as a single configuration of the environment). The different possible values of $x^j_i[t]$,$\bar X^j_i[t]$ and $Y^j[t]$ for the same (\textit{preparation of the}) many-body wave function $\Psi(x_1,x_2,...,x_N,y_1,y_2,...,y_M)$ introduce the inherent quantum randomness in any experiment. As such, if one is interested in ensemble average values, one can repeat the calculation for all environment configuration $Y^j[t]$ and particle distributions $x^j_i[t]$ and $\bar X^j_i[t]$. Typically, in electronics, this ensemble average of the current $I^j(t)$ over many experiments $j=1,...,\infty$ are interesting for evaluating DC values of the electrical current under ergodic assumptions. In the laboratory, a large time-average of the current $I^j(t)$ in a single $j$-experiment is performed. However, if one is interested in noise or time-correlations of the current at different times, $I^j(t_1)$ and $I^j(t_2)$, then the access to the individual experiment offered by the BCWF is very relevant.

Finally, we mention which are the computational advantages of this simulation framework. It is a microscopic description of the transport in the sense that it provides an individual description for each electron inside the active region. It provides a rigorous estimation (a part from the approximations for $\mathcal{A}^j(x_i,t)$ and $\mathcal{B}^j(x_i,t)$) to the quantum dynamics of electrons in the active region (open quantum system) for Markovian and non-Markovian systems. It is a versatile approach in the sense that it can simulate many different scenarios, from steady-state DC to transient and AC, including the fluctuations of the current (noise).  Notice that $I^j(t)$ in \eqref{ramo} includes the particle and displacement current, even at THz frequencies, when multi-time measurements are implicit. In this sense, we argue that the amount of information that this simulator framework can provide in the quantum regime is comparable to the predicting capabilities of the traditional Monte Carlo solution of the Boltzmann transport equation~\cite{jacoboni1983monte} in the semi-classical regime.

\section{How do we select the single-particle pure states before and after the collision?}
\label{s3}

To discuss how electron-photon scattering can be included in this simulation framework, we provide, first, an exact computation of the interaction between a single electron and a single photon in a closed system in terms of BCWF and Bohmian trajectories and, second, we provide some indications on how such interaction can be modeled (without the explicit consideration of the photon) in an open system, giving attention to how the initial and final states have to be modeled.

\subsection{Exact solution in a closed system}
\label{3.1}

\begin{figure}[h]
	\includegraphics[scale=0.08]{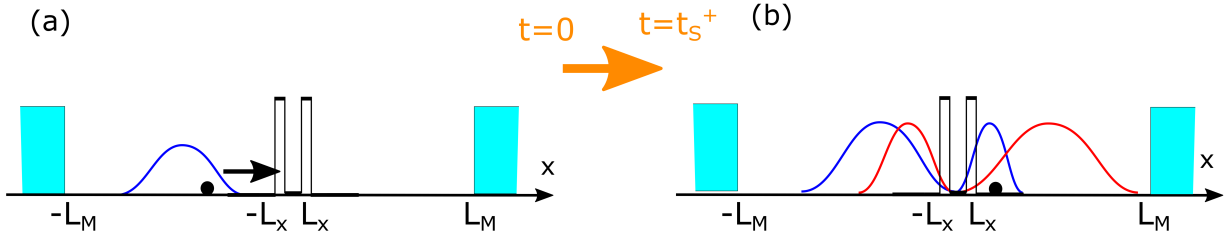}
	\caption{Schematic representation of the time evolution of the wave function for an electron impinging upon a double barrier region with electromagnetic radiation. In (a) and (b) we consider a cavity small enough so that the electromagnetic light does not radiate and no interaction with an environment degree of freedom outside the active region is included. Only the information on the electron degree of freedom $x$ and the internal degree of freedom of the light $q$ (not plotted) are relevant. The Bohmian position of the electron $X[t]$ is indicated as  a solid black circle. The $Q[t]$ trajectory of the electromagnetic field is not indicated. Notice that in (a) the initial electron wave function is $\psi_A(x,t=0)\neq0$ (blue curve for the electron) and $\psi_B(x,t=0)=0$, while in (b) we get $\psi_B(x,t=0) \neq 0$ due to spontaneous emission.}
	\label{Overlapping}
\end{figure}

The full quantum Hamiltonian $\hat H=\hat H_e+\hat H_\gamma +\hat H_I$ that describes light-matter interaction is given by the sum of the electron Hamiltonian $\hat H_e$, the electromagnetic field Hamiltonian $\hat H_\gamma$, and the electron-photon interaction Hamiltonian $\hat H_I$.  In particular, for a single electron in a semiconductor, the position representation for $\hat H_e$ (assuming a 1D system for the electron with degree of freedom $x$) is given by
\begin{equation}
H_e=-\frac{\hbar^2}{2m^*} \frac{\partial^2}{\partial x^2}+V(x),
\label{He}
\end{equation}
where $V(x)$ includes both internal and external electrostatic potential.  See the blue electron wave packet and the scalar potential $V(x)$ for a double barrier region of length $2L_x$ in the horizontal $x$-axis of Fig. \ref{Overlapping}(a).

We consider that the electromagnetic field is described by a single mode with angular frequency $\omega$ inside a closed cavity of length $2L_M$. See the cyan mirrors in the horizontal $x$-axis of Fig.~\ref{Overlapping}(a) and (b). The typical description of the electric field will be $E(x,t) \propto q\;cos (kx-\omega t)$ with the wave vector $k=2\pi/\lambda$ related to the angular frequency as $c=\omega /k$ with $c$ the speed of light. The variable $q$ represents the instantaneous amplitude of the electromagnetic field along the polarization vector. Under the assumption $L_M\gg L_x$ meaning that the wave-length for the electromagnetic wave ($\approx 500$ nm) is much larger than the active region ($\approx 20$ nm), we can neglect the spatial dependence $x$ of the electromagnetic field. Then, the Hamiltonian of the electromagnetic field in second quantization can be written as $\hat H_\gamma=\hbar \omega \left(1/2+ \hat a^\dagger \hat a\right)$. The relationship between the now quantized amplitude of the electric field $q$ and the creation $\hat a^\dagger$ and annihilation operators $\hat a$ is given by
\begin{equation}
\label{posrep}
\hat a=\sqrt{\frac{\omega}{2\hbar}} \left ( q+\frac{\hbar}{\omega}\frac {\partial }{\partial q}\right)\;\;\;\;,\;\;\;\hat a^\dagger=\sqrt{\frac{\omega}{2\hbar}} \left ( q-\frac{\hbar}{\omega}\frac {\partial }{\partial q}\right).
\end{equation}
Then, the q-representation of $\hat H_\gamma$ is
\begin{eqnarray}
H_\gamma =-\frac{\hbar^2}{2} \frac{\partial ^2 }{\partial q^2}+\frac{\omega^2}{2}q^2,
\end{eqnarray}
where the electromagnetic vacuum state with zero photons $|0\rangle$ solution of $\hat H_\gamma$ corresponds to the ground state of a harmonic oscillator $\psi_0(q)=\langle q|0\rangle$, while the state solution of $\hat H_\gamma$  with one photon corresponds to the first excited state of an harmonic oscillator  $\psi_1(q)=\langle q |\hat a^\dagger |0\rangle$.

\end{multicols}
\begin{multicols}{2}

The interaction Hamiltonian in the dipole approximation can be written as $\hat H_I=-e\hat x \hat E$,  where $e$ is the (unsigned) electron charge and the electrical field operator is given by $\hat E=\epsilon \left(\hat a+ \hat a^\dagger \right)$, with $\epsilon$ the strength of the electric field, or explicitly as
\begin{eqnarray}
H_I =\alpha' x q
\end{eqnarray}
where $\alpha'$, which depends on $\epsilon$ and other parameters of the cavity, controls the strength of the light-matter interaction. Finally, the wave function $\Psi(x,q,t)$ that describes the quantum nature of electrons and the electromagnetic field simultaneously in the $q$-representation is solution of the following two-dimensional Schr\"odinger equation,
\begin{eqnarray}
\label{scho}
i \hbar \frac{\partial \Psi(x,q,t)}{\partial t} =& -&\frac{\hbar^2}{2 m} \frac{\partial ^2 \Psi(x,q,t)}{\partial x^2}+V(x)\Psi(x,q,t) \nonumber\\
&-&\frac{\hbar^2}{2} \frac{\partial ^2 \Psi(x,q,t)}{\partial q^2}+\frac{\omega^2}{2}q^2\Psi(x,q,t)\nonumber\\
&+&\alpha' x q \Psi(x,q,t).
\end{eqnarray}
To simplify our discussion of emission and absorption of a photon by an electron, let us assume that only the zero photon state,  $\psi_0(q)=\langle q|0\rangle=\langle q|\psi_0\rangle$, and the one photon state,  $\psi_1(q)=\langle q |\hat a^\dagger |0\rangle=\langle q|\psi_1\rangle$, are relevant in our active region. Notice that we are discussing the interaction of a single electron with a single photon. Then, we can rewrite the wave function $\Psi(x,q,t)$ solution of \eqref{scho} as
\begin{eqnarray}
\label{super}
\Psi(x,q,t) = \psi_A(x,t)\psi_0(q)+\psi_B(x,t)\psi_1(q),
\end{eqnarray}
with
\begin{eqnarray}
\label{def_electron_side}
\psi_A(x,t)=\int \; \psi_0^*(q)\Psi(x,q,t) dq\\
\psi_B(x,t)=\int \; \psi_1^*(q)\Psi(x,q,t) dq.
\end{eqnarray}
The equation of motion of $\psi_A(x,t)$ and $\psi_B(x,t)$ can be obtained  by introducing the definition \eqref{super} into \eqref{scho} and using the orthogonality of $\psi_0(q)$ and $\psi_1(q)$ as follows
\begin{eqnarray}
\label{schoA}
i \hbar \frac{\partial \psi_A(x,t)}{\partial t} =& -&\frac{\hbar^2}{2 m} \frac{\partial ^2 \psi_A(x,t)}{\partial x^2} \nonumber \\ &+& \left(V(x) + \frac{1}{2}\hbar \omega\right)  \psi_A(x,t)\nonumber \\ &+&\alpha x \psi_B(x,t),\\ \\
i \hbar \frac{\partial \psi_B(x,t)}{\partial t} =& -&\frac{\hbar^2}{2 m} \frac{\partial ^2 \psi_B(x,t)}{\partial x^2} \nonumber \\ &+& \left(V(x)  + \frac{3}{2}\hbar \omega \right) \psi_B(x,t)\nonumber \\ &+&\alpha x \psi_A(x,t),
\label{schoB}	
\end{eqnarray}
where we have defined $\alpha=\alpha' \int \psi_0(q) q \psi_1(q) dq$ and we have used $\int \psi_0(q) q \psi_0(q)\;dq=\int \psi_1(q) q \psi_1(q)\;dq=0$.

\begin{figure}[H]
	\begin{minipage}{\linewidth}
		\centering
		{ \hspace*{0.8cm}
		\includegraphics[scale=0.2]{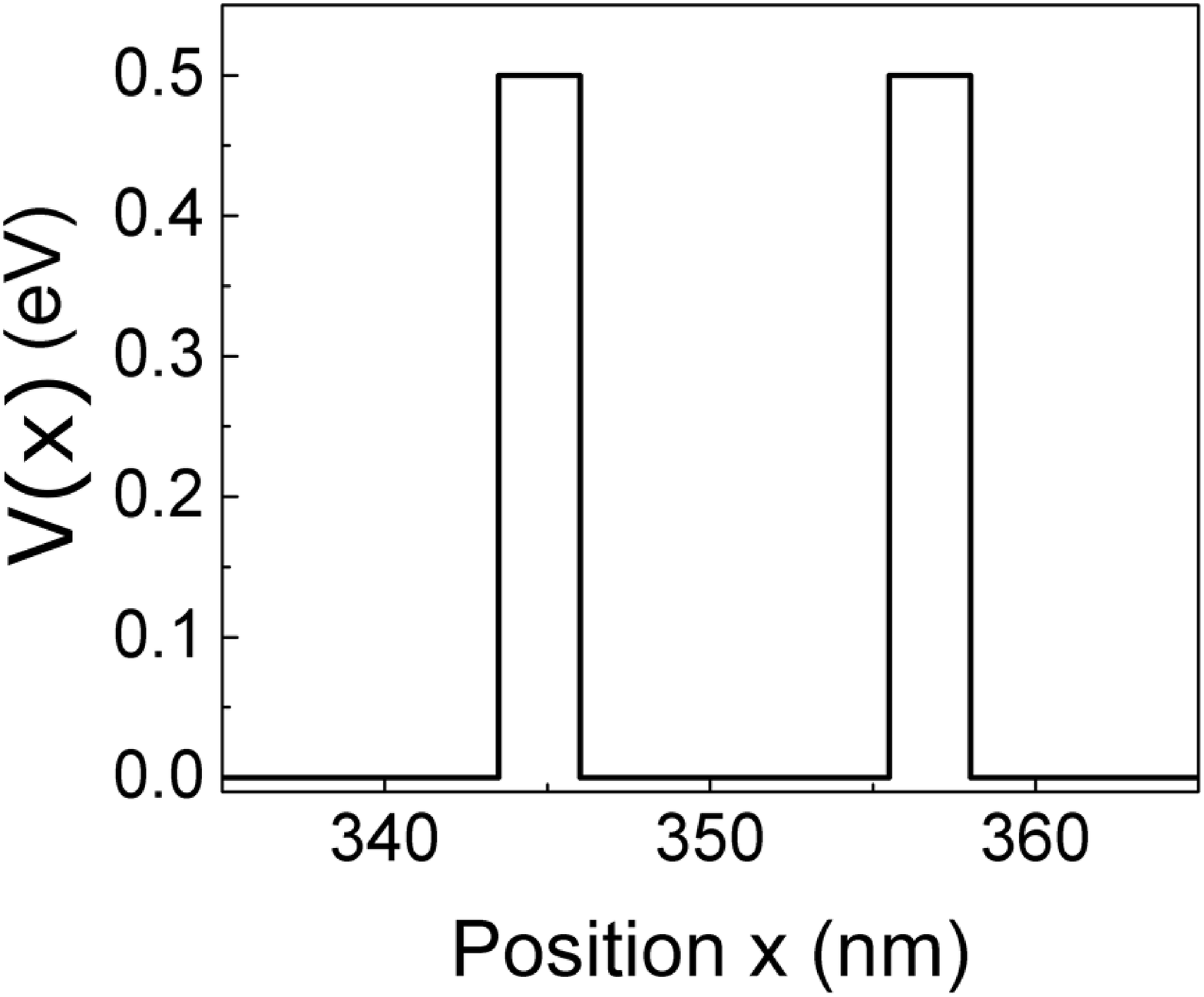}}
		{  \includegraphics[scale=0.28]{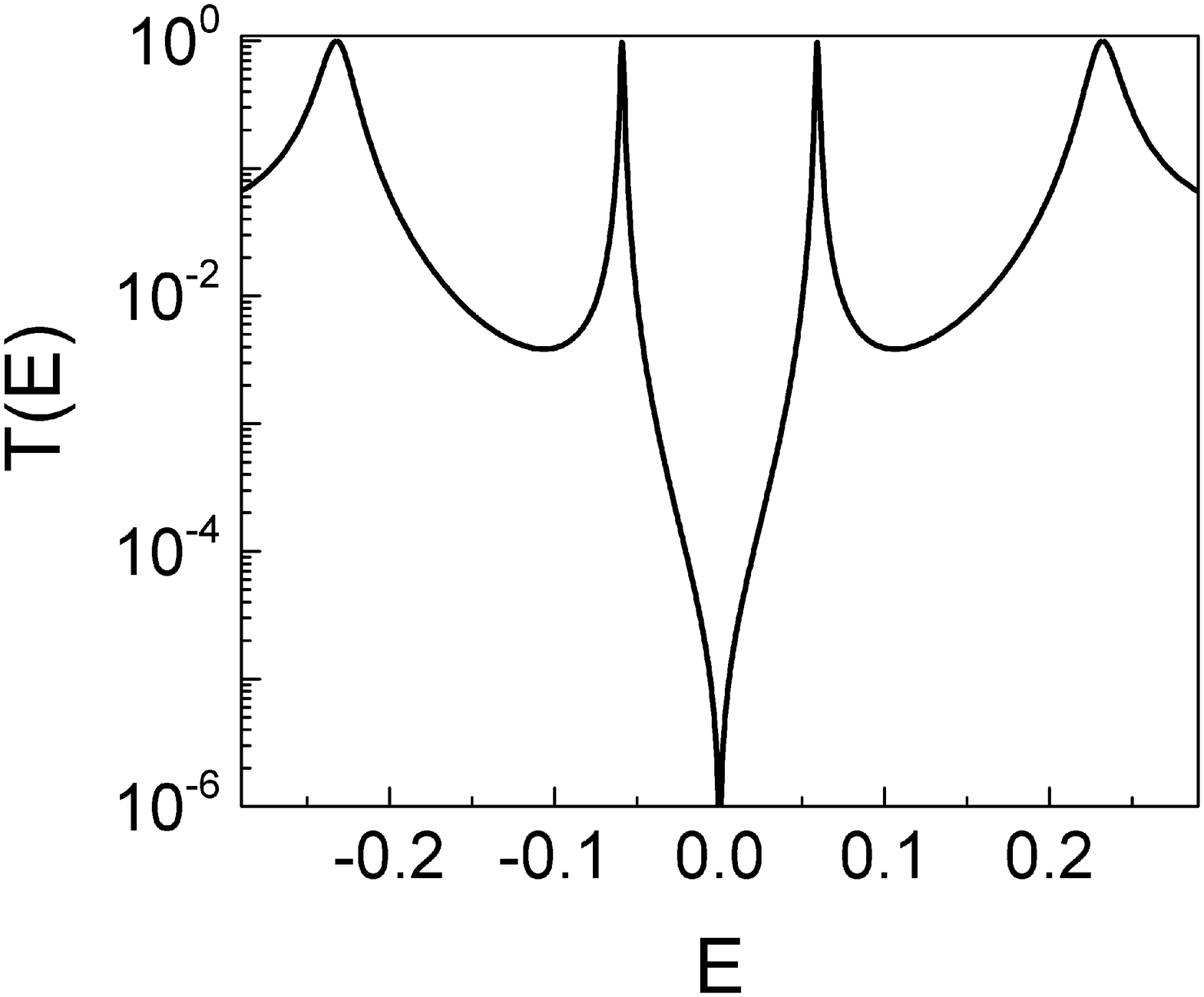}}
	\end{minipage}
	\caption{(a) Band structure and (b) Transmission coefficient of a GaAs/AlGaAs RTD device with 10 nm well width. Positive energies means electrons injected from the left and negative energies electrons injected from the right. }
	\label{gig1}
\end{figure}

We simulate now an initial electron impinging on a double barrier with a potential energy $V(x)$ as shown in Fig.~\ref{gig1} (a). It corresponds to the conduction band of a typical resonant tunnelling diode (RTD) with a 10 nm-well width, barrier thickness of 2 nm, and barrier height of 0.5 eV. In Fig.~\ref{gig1} (b), the transmission coefficient of the double barrier is plotted showing two resonant energies inside the well at  $E_1=0.058$~eV and $E_2=0.23$~eV. The positive energies correspond to energy eigenstates impinging from the left and negative energies from the right. 

At the initial time, we assume that there are no photons inside the active region. In other words, the (vacuum) electromagnetic field is given by an amplitude $q$ with probability $|\psi_0(q)|^2$. Thus, we define $\psi_A(x,0)$ as a Gaussian wave packet outside of the barrier region with a central energy equal to the second resonant level of the double barrier $E_2$ and a spatial dispersion of $30$ nm, as seen  in the blue wave packet in the $x$-axis of Fig. \ref{Overlapping}(a), and  $\psi_B(x,0)=0$. Thus, the initial electron-photon wave function in expression \eqref{super} is given only by $\Psi(x,q,t) = \psi_A(x,t)\psi_0(q)$. When solving \eqref{schoA} and \eqref{schoB}, together, with $\alpha=2.5\cdot 10^{7} eV/m$ and $\omega=(E_2-E_1)/\hbar$ we obtain that $\psi_B(x,t) \neq 0$ so that the global wave function in \eqref{super} becomes $\Psi(x,q,t) = \psi_A(x,t)\psi_0(q)+\psi_B(x,t)\psi_1(q)$. This process of spontaneous emission cannot be understood without the quantization of the electromagnetic field done in \eqref{scho}.

Next, to compute how much probability inside the well can be assigned to $\psi_A(x,t)$ and $\psi_B(x,t)$, at each resonant level, we define
\begin{eqnarray}
P_{A,1}(t)=\frac{1}{N}\int_{0}^{\frac{E_1+E_2}{2}} |c_A(E,t)|^2 dE,\\
P_{A,2}(t)=\frac{1}{N}\int_{\frac{E_1+E_2}{2}}^{\infty} |c_A(E,t)|^2 dE, \nonumber 
\end{eqnarray}
with
\begin{equation}
c(E,t)=\int_{-L_x}^{L_x} \psi(x,t) \phi_E^*(x) dx,
\label{localsuper}
\end{equation}
The subindex $A$ in $c(E,t)$ and $\psi(x,t)$ is assumed in \eqref{localsuper}. We omit it to use later, in the numerical results, the same expression for a general BCWF. The functions $\phi_E(x)$ are the energy eigenstates of the electron Hamiltonian $H_{e}$ in \eqref{He}. Notice that we are only interested in the probability inside the barrier region with limits given by $x=\pm L_x$. Identical definitions can be provided for $P_{B,1}(t)$ and $P_{B,2}(t)$ with the normalization constant $N$ ensuring that $P_{A,1}(t)+P_{A,2}(t)+P_{B,1}(t)+P_{B,2}(t)=1$.

\begin{figure}[H]
	{  \includegraphics[scale=0.3]{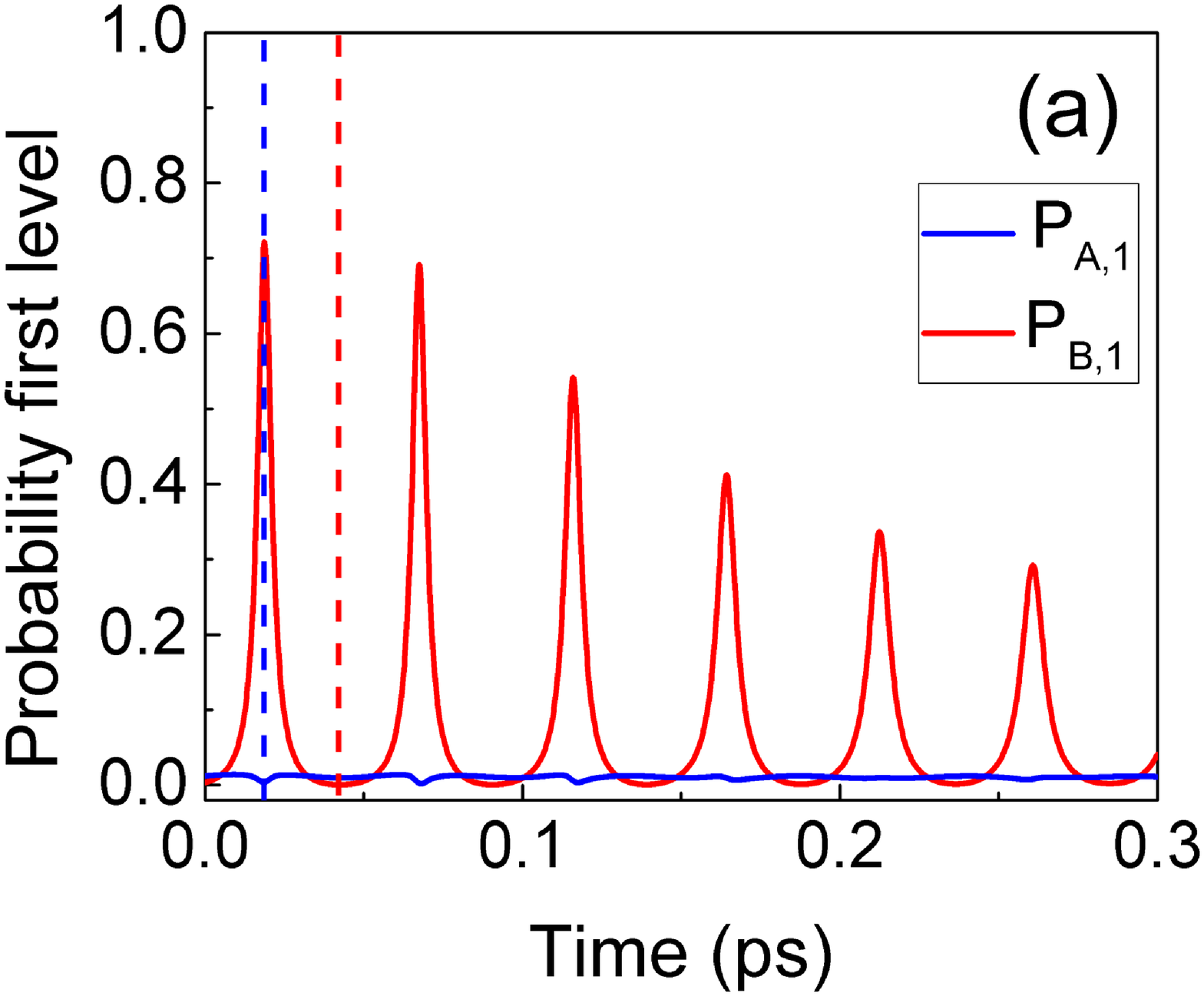}}
	\hspace*{-0.3cm}
	{ \includegraphics[scale=0.3]{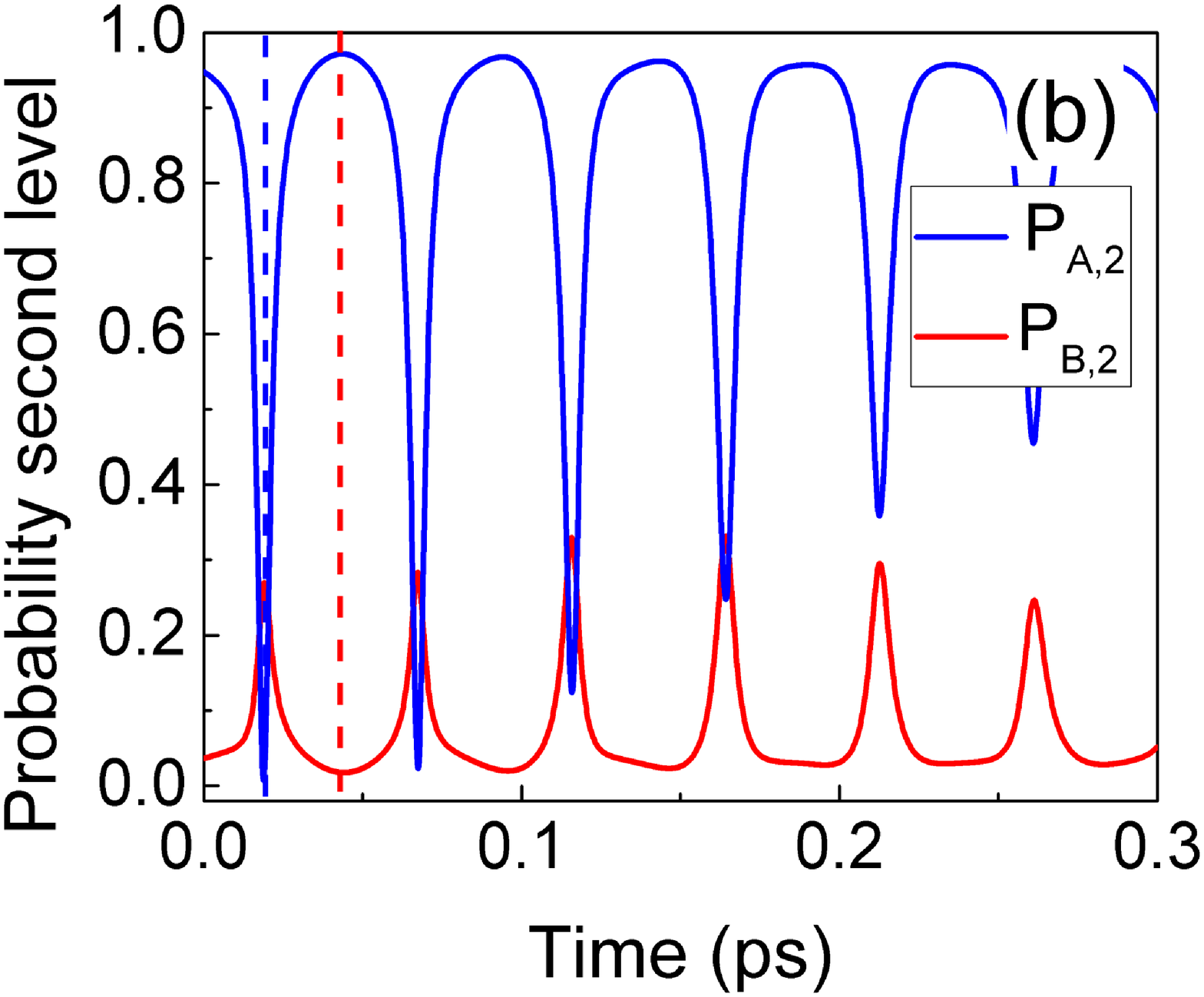}}
	\caption{Evolution of the $P_{A,1}$, $P_{A,2}$, $P_{B,1}$, $P_{B,2}$ for the first (a) and second (b) eigenstates of the quantum well described in Fig.\ref{gig1}, when the BCWF is injected in the second  eigenstate of the quantum well.}
	\label{comparison_FP_neg}
\end{figure}

In Fig.~\ref{comparison_FP_neg} we plot $P_{A,1}(t)$, $P_{A,2}(t)$, $P_{B,1}(t)$ and $P_{B,2}(t)$, show the typical Rabi oscillation. The initial value $P_{A,2}(0)\equiv 1$ in Fig. \ref{comparison_FP_neg} indicates an electron injected with a central energy equal to the second eigenvalue of the well without photons. The vertical dashed lines in Fig~\ref{comparison_FP_neg} indicates two times where the system passes from one electron in the first level and one photon $P_{A,2} \approx 0$ and $P_{B,1} \approx 1$ (blue dashed line), to one electron in the second level and zero photons $P_{A,2} \approx 1$ and $P_{B,1}\approx0$ (red dashed line).

\begin{figure}[H]
	{  \includegraphics[scale=0.24]{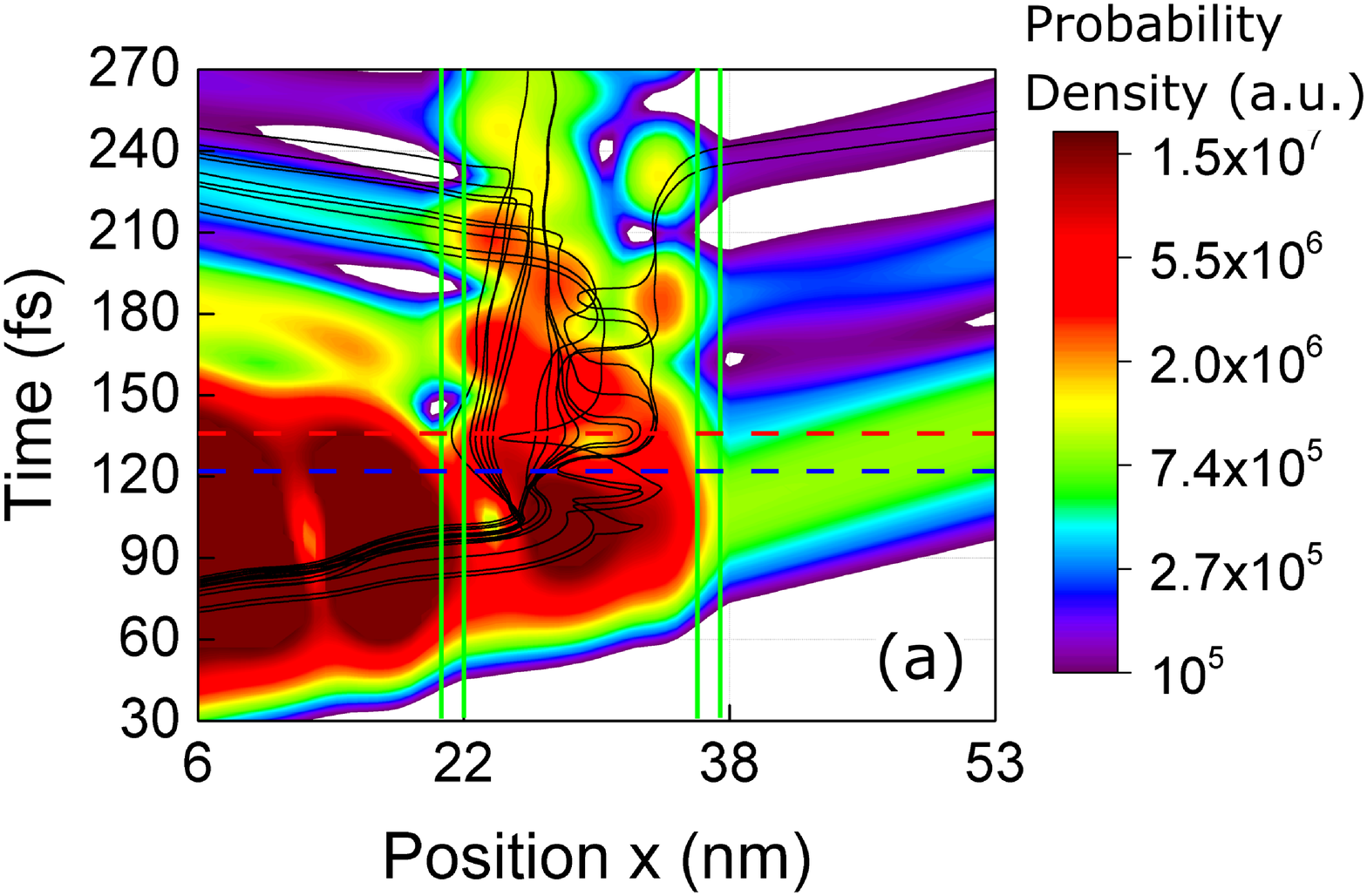}}
	\hspace*{-0.2cm}
	{ \includegraphics[scale=0.23]{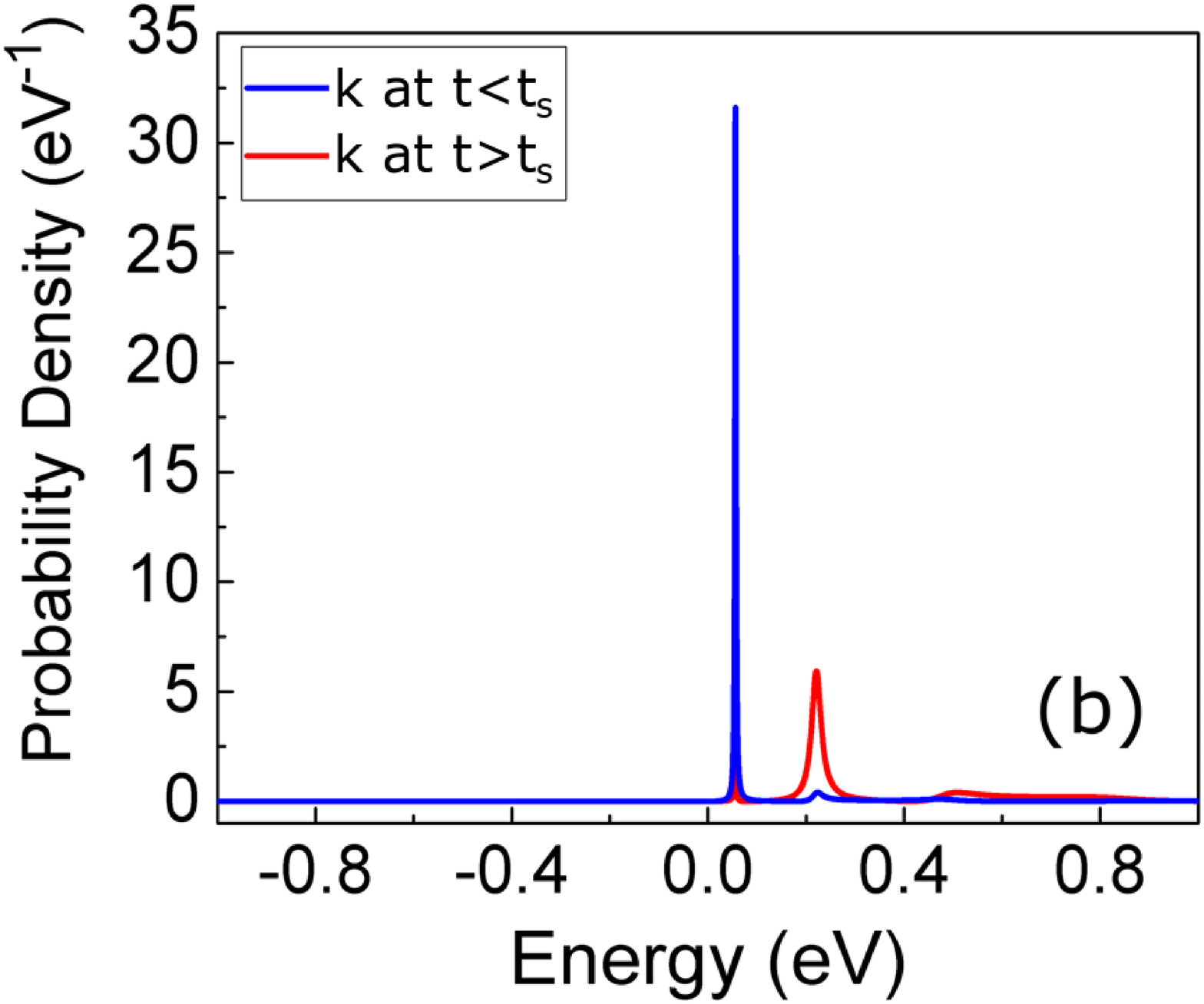}}
	\caption{(a) Evolution of $P_e(x,t)$ for the electron interacting with the RTD device described in Fig. \ref{gig1}, while emitting and absorbing electromagnetic radiation. The solid black lines show Bohmian trajectories $X^j[t]$ for a selected set of experiments. The green vertical lines indicate the position of the potential barriers. (b) Probability distribution of the Hamiltonian eigenstates for the BCWF given by $\psi(x,t)=\Psi(x,Q^j[t],t) = \psi_A(x,t)\psi_0(Q^j[t])+\psi_B(x,t)\psi_1(Q^j[t])$ for a selected trajectory $Q^j[t]$ at two different times indicated by the horizontal dashed lines in (a). We define the scattering time $t_s$ as the time of the blue horizontal dashed line.}
	\label{evolution_x_exact_model}
\end{figure}

From the whole wave function $\Psi(x,q,t) = \psi_A(x,t)\psi_0(q)+\psi_B(x,t)\psi_1(q)$ we can compute the probability presence in the $x$-space as:
\begin{equation}
P_e(x,t)=\int dq \Psi(x,q,t)=|\psi_A(x,t)|^2+|\psi_B(x,t)|^2
\label{P_e(x)}
\end{equation} 
In Fig. \ref{evolution_x_exact_model}(a) we have represented the evolution of $P_e(x,t)$ computed from \eqref{P_e(x)} as a function of time together with some selected trajectories $X^j[t]$. Such trajectories $X^j[t]$ are computed from the guiding total wave function $\Psi(x,q,t) = \psi_A(x,t)\psi_0(q)+\psi_B(x,t)\psi_1(q)$ together with the trajectories $Q^j[t]$ belonging to the electromagnetic degree of freedom $q$ following the velocities defined in \eqref{velo} for the same simulation presented before. The evolution shows qualitatively the alternate transition from the first to the second eigenstate of the quantum well. The Bohmian trajectories follow this evolution, since they alternatively move from the side to the center of the quantum well. This evolution suggests a type of Rabi oscillations where the electron emits a photon into a single-mode electromagnetic cavity and then reabsorbs it.
The evolution of $P_e(x,t)$ inside the well shows qualitatively the alternate transition from one maximum  (first eigenstate) to two maxima (second eigenstate). The Bohmian trajectories show a velocity close to zero when each eigenstate is well-defined, and a large velocity during the transitions between the two eigenstates. All this dynamical information is in agreement with the physics of the Rabi oscillations depicted in Fig. \ref{gig1} where the electron emits a photon into a single-mode electromagnetic cavity and then reabsorbs it. As a technical detail, we mention that as expected Bohmian trajectories do not cross in the $x-q$ space (not plotted), but they cross in the subspace $x$ of Fig. \ref{evolution_x_exact_model}(a). In addition, one can expect some chaotic behavior in 2D systems \cite{Tzemos1,Tzemos2} that is not present in the 1D system that is shown in the Fig. \ref{evolution_x_exact_model}(a).

In Fig. \ref{evolution_x_exact_model}(b) we plot the probability of the energy states $|c(E,t)|^2$ given by Eq.~\eqref{localsuper} at the two times indicated by horizontal read and blue dashed lines in Fig. \ref{evolution_x_exact_model}(a) that corresponds to the vertical dashed lines in Fig~\ref{comparison_FP_neg}. The BCWF in  Eq.~\eqref{localsuper} has been defined as  $\psi(x,t)=\Psi(x,Q^j[t],t) = \psi_A(x,t)\psi_0(Q^j[t])+\psi_B(x,t)\psi_1(Q^j[t])$ for a selected trajectory $Q^j[t]$ of the $j$-experiment. Notice that such definition of the BCWF corresponds in this case to $\psi(x,t) \approx \psi_B(x,t)$ for the blue wave packet, while the red wave packet corresponds to $\psi(x,t) \approx \psi_A(x,t)$, because of the values of $P_{A,2}$ and $P_{B,1}$ when Fig. \ref{evolution_x_exact_model}(b) is estimated.

\begin{figure}[H]
	{  \includegraphics[scale=0.3]{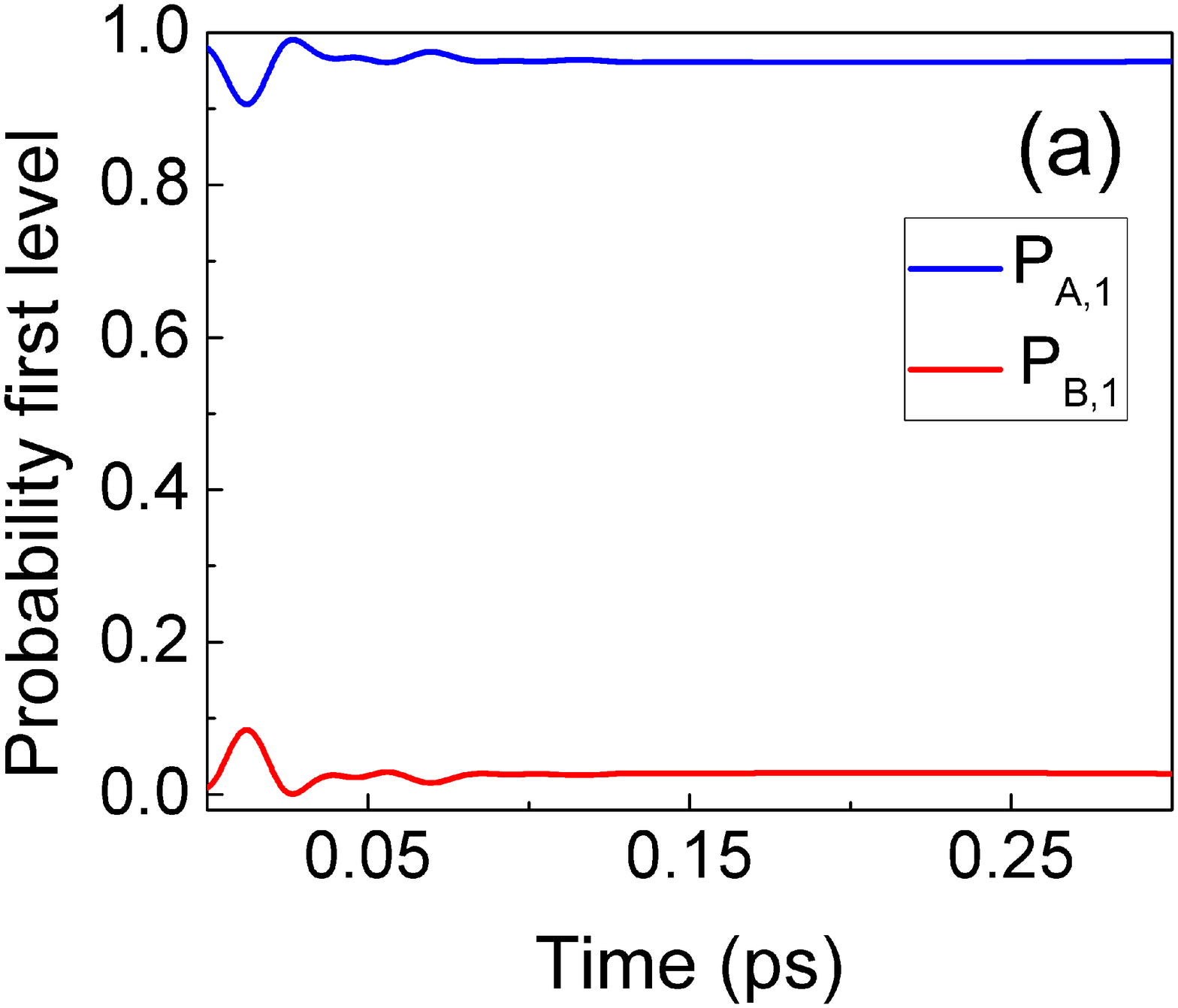}}
	\hspace*{-0.3cm}
	{ \includegraphics[scale=0.3]{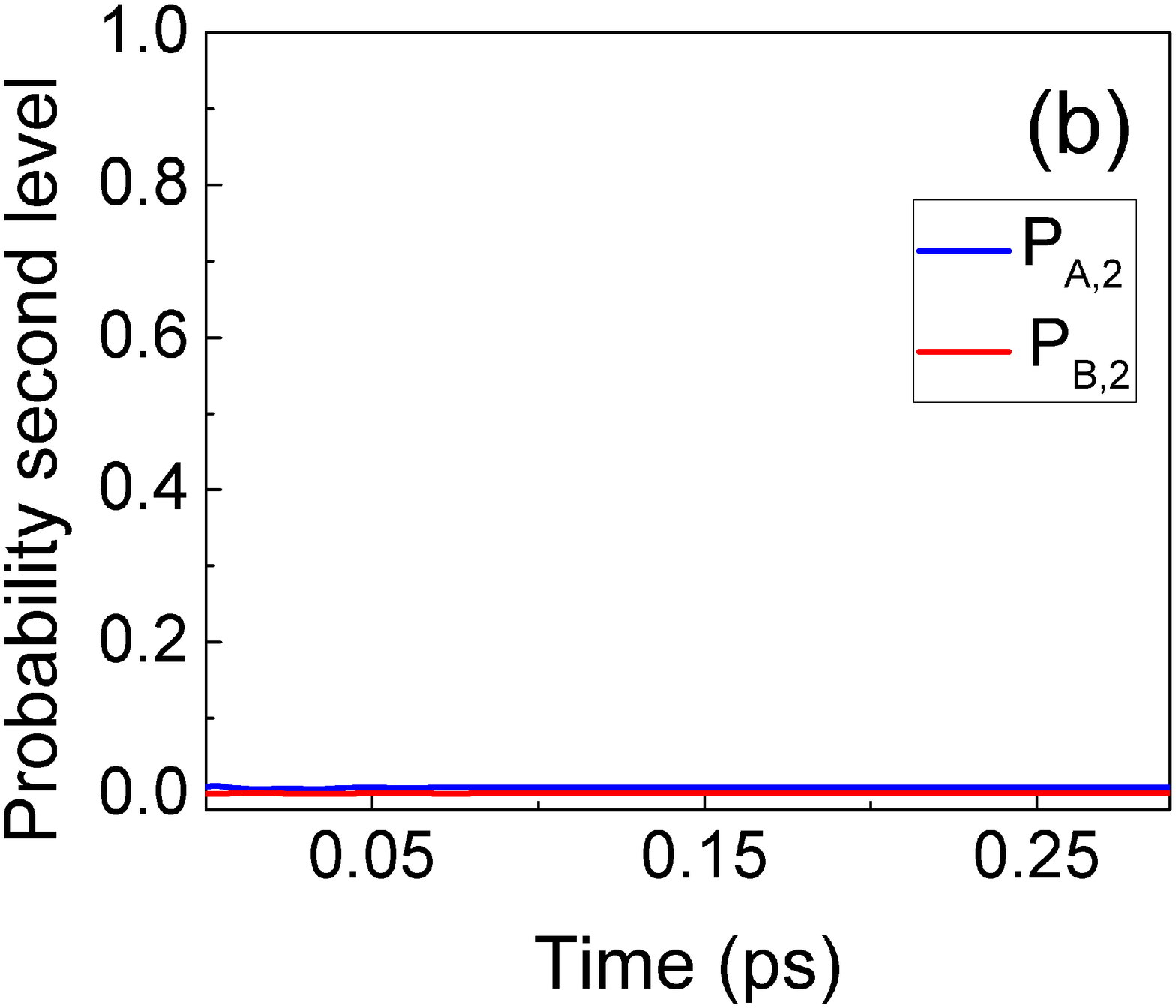}}
	\caption{Evolution of the $P_{A,1}$, $P_{A,2}$, $P_{B,1}$, $P_{B,2}$ for the first (a) and second (b) eigenstates of the quantum well when the initial electron is injected in the first resonant level of the quantum well. Because of the conservation of energy, no matter-light interaction is possible.}
	\label{evolution_prob_1st_level_inj}
\end{figure}

As expected, the fact that the conservation of the total energy has to be satisfied from \eqref{scho} has important consequences on the type of electron-photon interaction allowed. We now repeat the simulation when the electron (with  no photon) is injected with a central energy corresponding to the first resonant level. No electron transition (or spontaneous emission) takes place, giving $\psi_B(x,t)\approx 0$ because the initial energy $E_1+\hbar \omega/2$ cannot be converted into a much higher final energy $E_2 + 3 \hbar \omega/2$. The result is shown in Fig. \ref{evolution_prob_1st_level_inj}.

\begin{figure}[H]
	{  \includegraphics[scale=0.3]{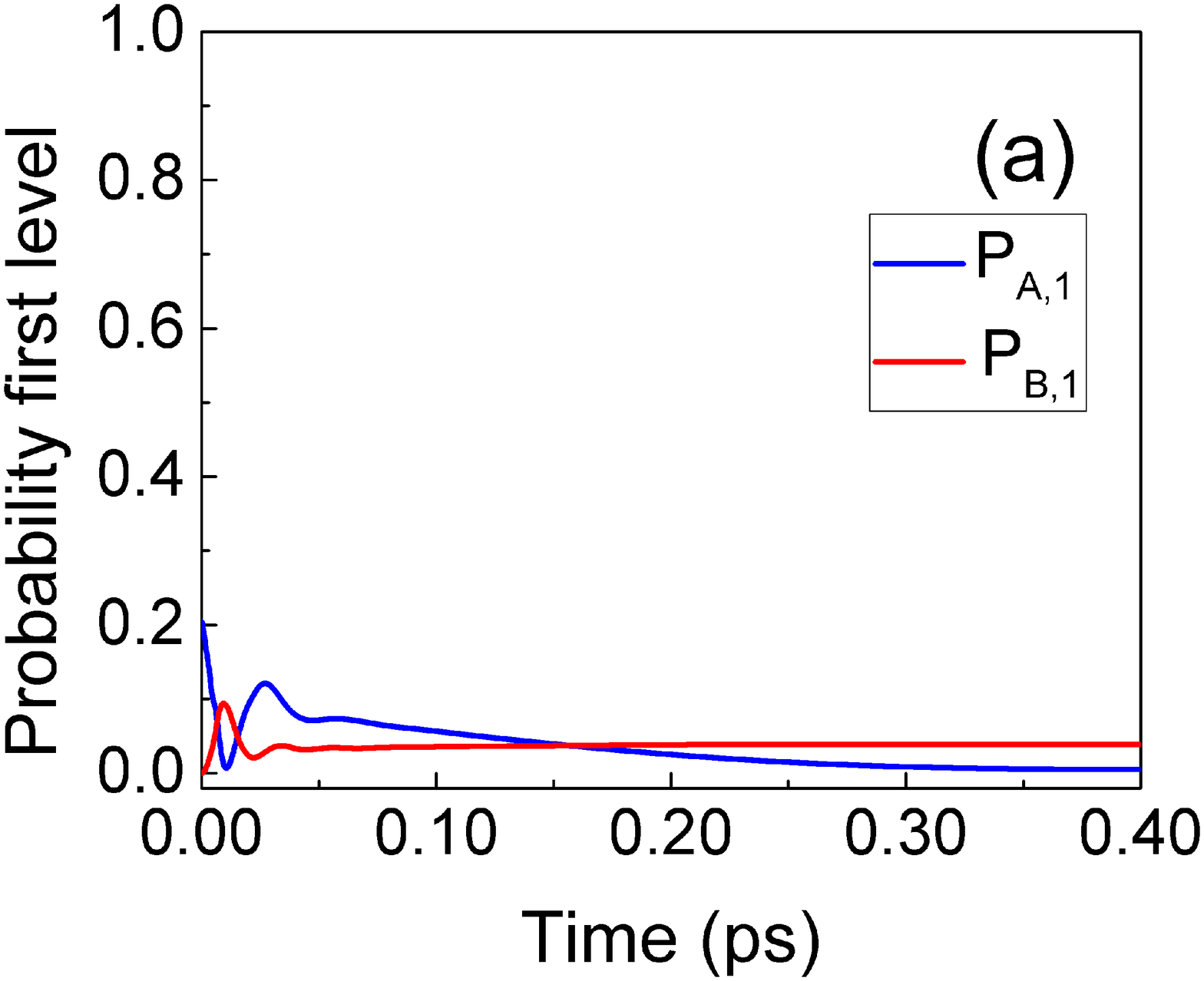}}
	\hspace*{-0.3cm}
	{ \includegraphics[scale=0.3]{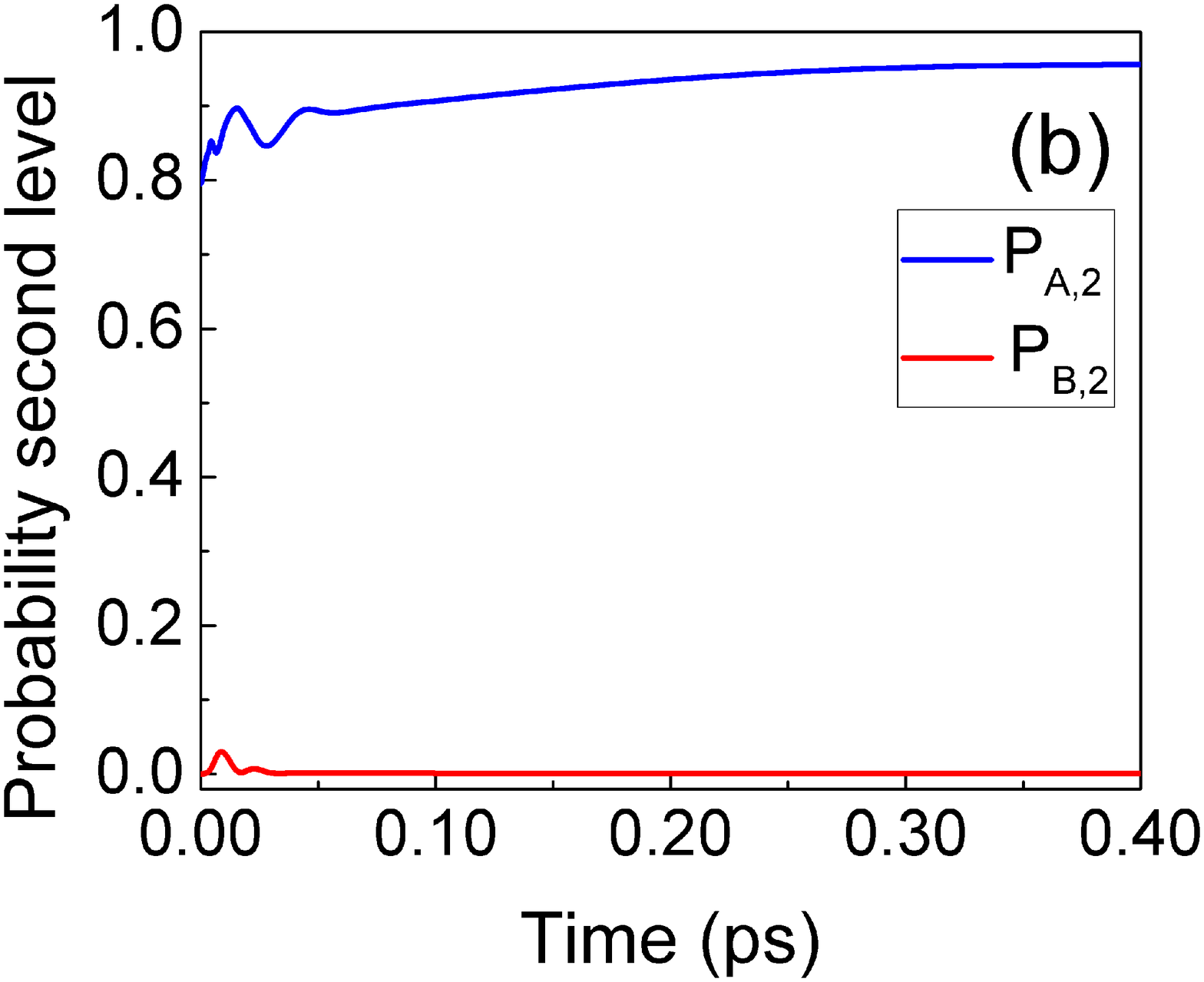}}
	\caption{Evolution of the $P_{A,1}$, $P_{A,2}$, $P_{B,1}$, $P_{B,2}$ for the first (a) and second (b) eigenstates of the quantum well when the BCWF is injected in the second eigenstate of the quantum well and $\hbar \omega = 0.26$ eV. Because of the conservation of energy, no matter-light interaction is possible.}
	\label{evolution_prob_noconservation}
\end{figure}

We now repeat the same simulation done in Fig \ref{comparison_FP_neg}, where the initial electron had mean energy equal to the second eigenvalue of the well, $E_2$, but considering a new photon energy $\hbar \omega = 0.26$ eV much larger than $E_2-E_1=0.172$ eV. In this case no light-matter interaction takes place since it would imply a violation of the conservation of whole energy. The initial energy $E_2+\hbar \omega/2$  do not coincide with a possible final energy $E_1 + 3 \hbar \omega/2$. This simulation is shown in Fig. \ref{evolution_prob_noconservation}.

\subsection{Approximate solution with BCWF for an open system}
\label{3.2}

In the previous subsection we discussed the interaction of a single electron with a single photon in a closed system. Here, we discuss how such interaction can be generalized to include the possibility to detect a photon at a position $y$, far from the active region. 

The proper simulation of such scenario as a closed system is far from the scope of the present paper. Apart from considering the detector outside of the active region as a new electron with degree of freedom $y$, the transition of the electromagnetic energy from the active region to the environment will require an electromagnetic field with an arbitrary shape different from the one considered in the previous section. A Fourier transform of such arbitrary electromagnetic field will imply dealing with several components $E(x,t) \propto q\;cos (kx-\omega t)$ at different frequencies. In any case, without an explicit solution of such problem, only from the conservation of energy, we can anticipate what will be the expected behavior of the whole system.

The process of spontaneous emission of a photon inside the active region, and its posterior detection far from the active region,  can be anticipated as:
\begin{itemize}
	\item At the initial time, $t=0$, we consider an electron in the active region, with degree of freedom $x$ with a central energy $E_2$, linked to zero photons wave function $\psi_0(q)$, plus another electron far from the active region,  with degree of freedom $y$ and energy $E_{ext}$, linked to zero photons $\psi_0(q)$. At this initial time, thus, the total energy involved in such scenario is $E_2+\hbar\omega/2$, in the active region, plus the energy $E_{ext}+\hbar\omega/2$ outside.
	\item At the intermediate time, we consider that a spontaneous emission of a photon happens inside the active region. As seen in Fig. \ref{comparison_FP_neg}, such internal process ensure energy conservation. So that the new photon inside the active region implies a change of energy there,  $E_2+\hbar\omega/2 \to E_1+3\hbar\omega/2$, while the energy outside of the active region remains the same as before, $E_{ext}+\hbar\omega/2$. The total energy is the same as the initial one. 
	\item At the final time $t$, we detect a photon at the position $y$ far from the active region. Thus, the electron at $y$ is now linked to one photon wave function $\psi_1(q)$ which implies an increment of the energy of $\hbar \omega$ far from the active region, $E_{ext}+\hbar\omega/2 \to E_{ext}+3\hbar\omega/2$. The conservation of the total energy implies that the same amount of energy is eliminated in the active region when the photon is leaving it, $E_1+3\hbar\omega/2 \to E_1+\hbar\omega/2$. The electron in the active region will have a new energy $E_1$ linked to the zero photons wave function $\psi_0(q)$. As we have seen in Fig. \ref{evolution_prob_1st_level_inj}, under such new energy conditions in the active region, such electron will not be able any more to generate spontaneous emission inside the RTD.  Thus, the Rabi oscillations seen in Fig. \ref{comparison_FP_neg} for a closed system, will not be present when we assume that the photon is leaving the cavity.
\end{itemize}

In summary, using the BCWF to focus only on the description of the electron inside the active region as a single-particle pure state $\psi(x,t)$, we conclude that the spontaneous emission in the active region can be modeled by an initial BCWF $\psi(x,0)$ with central energy $E_2$ that changes to a final BCWF $\psi(x,t)$ with energy $E_1$. Such process will be allowed as far as the photon energy coincides with $E_2-E_1$. Identically, absorption in the active region can be modeled by an initial BCWF $\psi(x,0)$ with central energy $E_1$ that changes to a final BCWF $\psi(x,t)$ with energy $E_2$ with the photon energy given by $E_2-E_1$. The conservation of energy enables the photon absorption to be accompanied by a subsequent process of spontaneous emission, that returns the photon energy to the environment outside of the active region.  

\section{Implementation of the transition from pre- to post-selected BCWF}
\label{s4}

In this section we describe practical issues on how such type of transitions between initial and final states can be implemented in a transport simulator for real electron devices based on Bohmian mechanics. Some additional information about these types of simulators for quantum transport can be found in the BITLLES simulator\cite{BITLLES1}.

To implement the transition from pre- to post-selected BCWF, a definition of the initial $|i \rangle$ and the final $|f \rangle$ states is needed. Although the contacts do not allow us to perfectly \textit{prepare} the wave description of electron, we can have some reasonable arguments to anticipate some of its properties.  One option could be to deal with Hamiltonian eigenstates, which extend to infinite in both sides (left and right) of the device. Although these infinitely-extended states are useful tools to model (steady-state) DC transport properties of quantum devices, they are less useful to describe other device performances as, for example, the fluctuations of the electrical current due to the partition noise in a tunneling barrier. The initial electron, after impinging with the barrier, is either located at the left (reflection)  or at the right (transmission) of a barrier, but not at both sides of it. Such randomness (transmission of reflection) translates into current fluctuations. To model such fluctuations, a localized wave function seems appropriate to model electrons.  However, the wave function cannot have a very narrow localization in position since the Heisenberg uncertainty principle would lead to extremely large momentum and energy uncertainties (larger than thermal energies). Thus, a definition of an electron, deep inside the contact, as a Gaussian wave packet with well-defined central position and central energy seems reasonable. We add that such limited extension of the electron wave function can be related to the coherence length of the sample.

In classical mechanics, an electron having a well-defined energy is compatible with an electron having a well-defined momentum. However, this is not the case for quantum electrons.  As a general rule, two properties can be simultaneously well-defined if their operators commute. In our case, the energy (linked to the Hamiltonian operator $\hat H$) and the momentum (linked to the momentum operator $\hat p$) can be simultaneously defined when $[\hat H, \hat p]=0$. In the position representation, knowing that the Hamiltonian operator is the sum of the kinetic energy operator $(\hat p)^2/2m$, which obviously commutes with $\hat p$, plus the potential energy operator $\hat V$, momentum and energy are well-defined properties when
\begin{equation}
\label{commu_V_p}
\big[ H, -i\hbar\frac{\partial}{\partial_x} \big]=\big[ V(x), -i\hbar\frac{\partial}{\partial_x}  \big]=i\hbar\frac{\partial V(x)}{\partial_x}=0.
\end{equation}
Thus, only when dealing with flat potentials we can assume that a wave packet with a reasonable well-defined energy has also a reasonable well-defined momentum. This discussion seems relevant to transport models developed in phase-space (the Wigner distribution function), where information on only momenta and positions are available.

In the next two subsections, we discuss the implementation of the transition from a pre- to a post-selected BCWF when using well-defined energies (model A) or momenta (model B). In Sec. \ref{s5} we compare the numerical results of these two different implementations.

\subsection{Model A: change of the central energy}
\label{s4.1}

We consider an electron defined by a single-particle BCWF that at time $t_s$ undergoes a scattering event. We define $t_s^-=t_s-\Delta t_s$ as the time just before and $t_s^+=t_s+\Delta t_s$ as the time just after the scattering event. For simplicity, we consider $\Delta t_s \to 0$, but we have seen in Sec. \ref{s3} that such transition between initial and final BCWF takes a finite time because, from a conceptual point of view, it has to guarantee the continuity of the BCWF in space and time. The initial and final BCWFs are $\psi(x,t_s^-)$ and $\psi(x,t_s^+)$, which satisfy $\langle E(t_s^+) \rangle=\langle E(t_s^-) \rangle + E_\gamma$, with $E_\gamma$ the energy of a photon. Within the energy representation, the wave packet can be decomposed into a superposition of Hamiltonian eigenstates $\phi_E(x)$ of the electron $\hat H_e$ in \eqref{He} as 
\begin{equation}
\psi(x,t_{s^-})=\int dE \; a(E,t_s^-) \; \phi_E(x),
\end{equation}
with $a(E,t)=\int dx \;\psi(x,t) \; \phi_E^*(x) $. The central energy $\langle E(t_s^-) \rangle$ is
\begin{equation}
\langle E(t_s^-) \rangle=\int dE \; E \; |a(E,t_s^-)|^2,
\end{equation}
which can be increased to get the new central energy at $t_s^+$ as
\begin{eqnarray}
\langle E(t_s^+) \rangle&=&\langle E(t_s^-) \rangle+E_\gamma\nonumber \\ &=& \int dE \; (E+E_\gamma) \; |a(E,t_s^-)|^2\nonumber \\ &=& \int dE' \; E' \; |a({E'-E_\gamma},t_s^-)|^2\nonumber\\
&=&\int dE' \; E' \; |a'(E',t_s^+)|^2,
\end{eqnarray}
where we have defined $a'(E,t_s^+)=a(E-E_\gamma,t_s^-)$. Thus, the new wavepacket after the collision is
\begin{eqnarray}
\psi(x,t_{s^+})&=&\int dE \; a'(E,t_s^+) \; \phi_E(x)\nonumber \\ &=& \int dE \; a(E'-E_{\gamma},t_s^-) \; \phi_E(x).
\end{eqnarray}
This transition corresponds to absorption of energy by the electron. Emission can be identically modeled by using $\langle E(t_s^+) \rangle=\langle E(t_s^-) \rangle - E_\gamma$. If required, the \emph{technical} discontinuity between $\psi(x,t_{s^-})$ and $\psi(x,t_{s^+})$ can be solved by just assuming that the change of energy is produced in a finite time interval $\Delta t_s=N_{t_s} \Delta t$, with $\Delta t$ the time step of the simulation. Then, at each time step of the simulation, the change in the wave packet central energy is $E_\gamma/N_{t_s}$. A \textit{continuous} change of both energy and wave packet will be obtained as far as $\Delta t \to 0$.

This continuous evolution of the BCWF can be represented as a Schr\"odinger-like equation as explained in \cite{PRLxavier}. In any case, the shape of the Hamiltonian of this new Schr\"odinger equation describing the collision is subjected to the post-selection of the state.

\subsection{Model B: change of central momentum}
\label{s4.2}

In Ref. \cite{Enrique1} we explain how a change of momentum $p_\gamma$ in a wave packet in free space can be performed with a unitary Schr\"odinger equation.  That algorithm can be understood as a pre- and a post-selection of the initial BCWF, $\psi(x,t_s^-)$, and of the final BCWF, $\psi(x,t_s^+)$, respectively. At time $t_s^-$, the BCWF can be written as a superposition of momentum eigenstates $\phi_p(x)$ (which are a basis of the electron in the x space) as

\begin{equation}
\psi(x,t_{s^-})=\int dp \; b(p,t_{s^-}) \; \phi_p(x),
\end{equation}
with $b(p,t_{s^-})=\int dx \psi(x,t_{s^-}) \; \phi_p^*(x) $. The central momentum $\langle p(t_s^-) \rangle$ is
\begin{equation}
\langle p(t_s^-) \rangle=\int dp \; p \; |b(p,t_{s^-})|^2,
\end{equation}
which can be increased to get the new central momentum $\langle p(t_s^+) \rangle=\langle p(t_s^-) \rangle+p_\gamma$ at $t_s^+$ as 
\begin{eqnarray}
\langle p(t_s^+) \rangle&=&\langle p(t_s^-) \rangle+p_\gamma \nonumber \\ &=& \int dp \; (p+p_\gamma) \; |b(p,t_{s^-})|^2\nonumber \\ &=& \int dp' \; p' \; |b({p'-p_\gamma},t_{s^-})|^2\nonumber\\
&=&\int dp' \; p' \; |b(p',t_{s^+})|^2,
\end{eqnarray}
where we have defined $b(p,t_{s^+})=b(p-p_\gamma,t_{s^-})$. In this particular scenario, we know the explicit shape of the momentum eigenstates, $\phi_p(x)=1/\sqrt{2\pi}\exp(i p x /\hbar)$, so that
\begin{eqnarray}
\psi(x,t_s^+)&=&\int dp \; b(p,t_{s^+}) \; \phi_p(x) \nonumber \\ &=& \int dp \; b(p-p_\gamma,t_{s^-}) \; \phi_p(x)  \nonumber\\
&=& \int dp \int dx' \psi(x,t_{s^-}) \; \phi_{p-p_\gamma}^*(x') \phi_p(x)  \\
&=& \int dp \int dx \; \psi(x,t_{s^-}) \; \frac{1}{2\pi} e^{i p(x'-x)/\hbar} e^{i p_\gamma x' /\hbar} \nonumber \\ &=& e^{i p_\gamma x /\hbar} \psi(x,t_{s^-}).\nonumber
\end{eqnarray}
With the condition $\psi(x,t_s^+)= e^{i p_\gamma x /\hbar} \psi(x,t_{s^-})$, it can be easily demonstrated which one is the unitary equation satisfied by the BCWF. If we define $\psi'(x,t)$ as the wave function solution of the following Schr\"odinger equation, $i \hbar \frac{\partial \psi'(x,t)}{\partial t} =\frac{1}{2m^*} \left(-i\hbar \frac{\partial}{\partial_x}\right)^2\psi'(x,t) +V(x)\psi'(x,t)$, with initial condition at $t=t_s$ given by $\psi'(x,t_s)=\psi(x,t_s^+)$, then, the solution $\psi'(x,t)$ for $t>t_s$ is identical to the following Schr\"odinger equation, $i \hbar \frac{\partial \psi(x,t)}{\partial t} =\frac{1}{2m^*} \left(-i\hbar \frac{\partial}{\partial_x} +p_\gamma \right)^2 \psi(x,t) +V(x)\psi(x,t)$, for the original $\psi(x,t)$ and with its original initial condition for $t>t_s$. Finally, a single equation for $\psi(x,t)$ valid for all times is just
\begin{eqnarray}
i \hbar \frac{\partial \psi(x,t)}{\partial t} &=&\frac{1}{2m^*} \left(-i\hbar \frac{\partial}{\partial_x} +p_\gamma \Theta_{t_s} \right)^2 \psi(x,t) \nonumber \\ &+& V(x)\psi(x,t),
\label{schomomentum}
\end{eqnarray}
where $\Theta_{t_s}$ is a Heaviside function equal to $1$ for $t>t_s$ and zero otherwise. Thus, a description of the evolution of the wave function $\psi(x,t)$ during the collision process can be done from a unitary Schr\"odringer equation, where the momentum operator $-i\hbar \frac{\partial}{\partial_x}$ is changed for the new momentum operator $-i\hbar \frac{\partial}{\partial_x} +p_\gamma \Theta_{t_s}$, as indicated in \cite{Enrique1}. Notice that the probability presence of the scattered wave packet satisfies $|\psi(x,t_{s^+})|^2=|\psi(x,t_{s^-})|^2$ because only a global phase $e^{i p_\gamma x /\hbar}$ is added.

It is quite easy to see from \eqref{velo2} that the Bohmian velocity of the electron after the collision computed from $\psi(x,t_s^+)$ is just the old velocity computed from $\psi(x,t_s^-)$ plus $p_\gamma/m^*$,
\begin{eqnarray}
v_x^j[t_s^+]&=&\frac {1}{m^*}\frac{\partial s(x,t_s^+)}{\partial x}|_{x=X^j[t]} \nonumber \\ &=&  \frac{1}{m^*}\frac{\partial s(x,t_s^-)}{\partial x}|_{x=X^j[t]}+p_\gamma/m^*.
\label{extravelo}
\end{eqnarray}
The collision increases the velocity of the electron by the same amount that we add in \eqref{schomomentum}. Unfortunately, as discussed at the beginning of the section, a global mechanism of scattering valid for scenarios with potential barriers requires dealing with change of the energy as presented in Model A (not with change of the momentum as presented in Model B).

\section{Numerical results}
\label{s5}
We present now the numerical results of our two models for the transition between initial and final single-particle BCWF, as explained in the previous section. We first study electron-photon collisions in free space, when energy and momentum operators commute, and then electron-photon collisions in a scenario with a double barrier potential profile, when energy and momentum operators do not commute. This last case will be compared with numerical results of the exact model presented in Sec. \ref{s3}, and used to verify the physical soundness of the two models.

\subsection{Collisions in flat potentials}
\label{s5.1}
In this section, we study the interaction of an electron and a photon in free space. The electron evolves in a flat potential. We consider the absorption of a photon by an electron.  In flat potential, the momentum and energy conservation  is ensured during the collision. Thus, since the momentum of the photon is negligible, in this Sec. \ref{s5.1}, we assume that the electron is interacting with a phonon and a photon. The phonon will not be needed in Sec. \ref{s5.2}. We consider that the final BCWF will be modeled by a final electron (post-selected state) with an energy increase of $\hbar \omega$ ($E_\gamma>0$) plus the corresponding increase of momentum (provided by the phonon) with respect to the initial electron energy (pre-selected state).
\end{multicols}
\begin{figure}[H]
	\begin{minipage}{\linewidth}
		\hspace*{-0.3cm}
		\centering
		{  \includegraphics[scale=0.28]{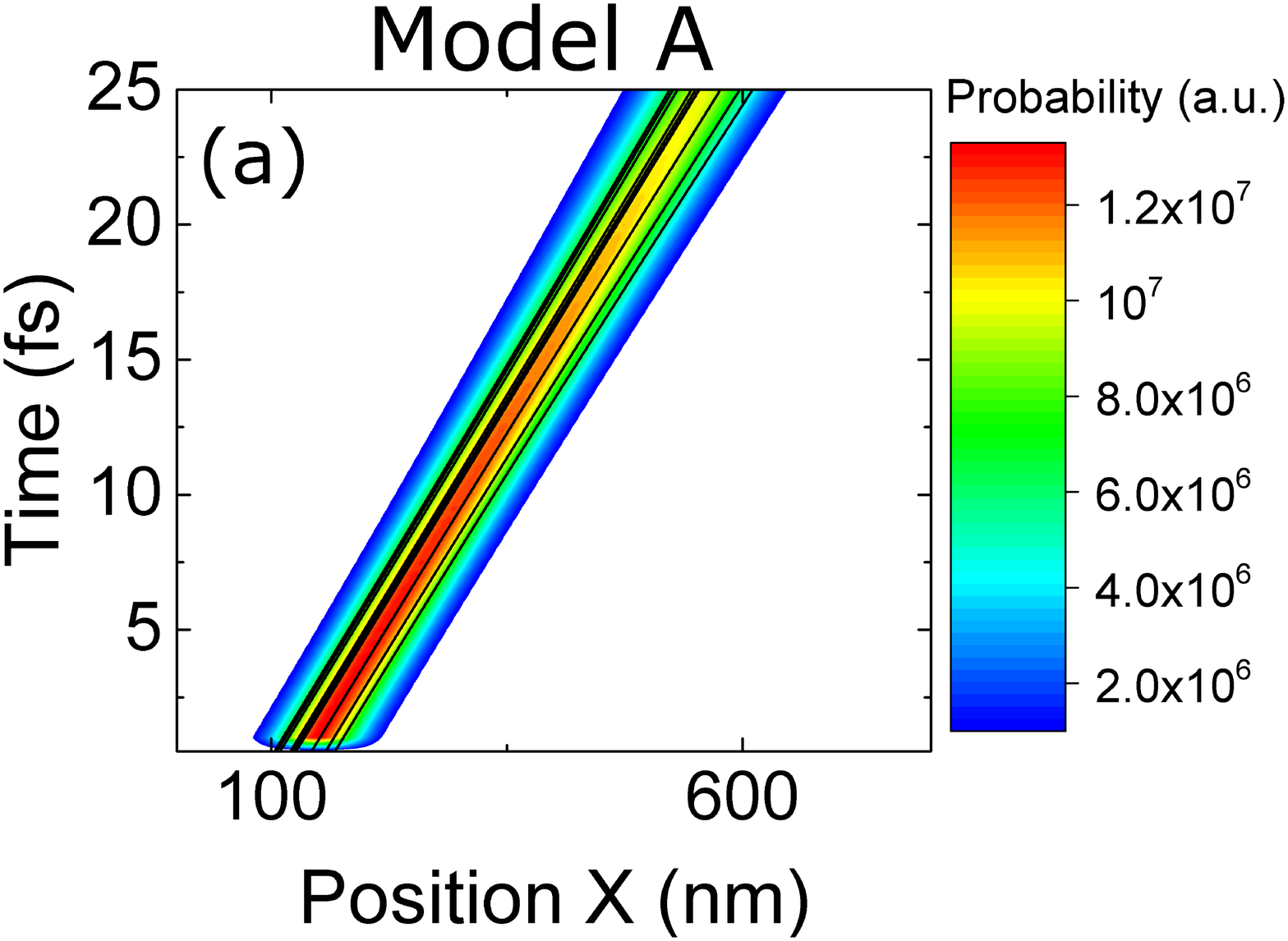}}
		\hspace*{-2.65cm}
		{  \includegraphics[scale=0.28]{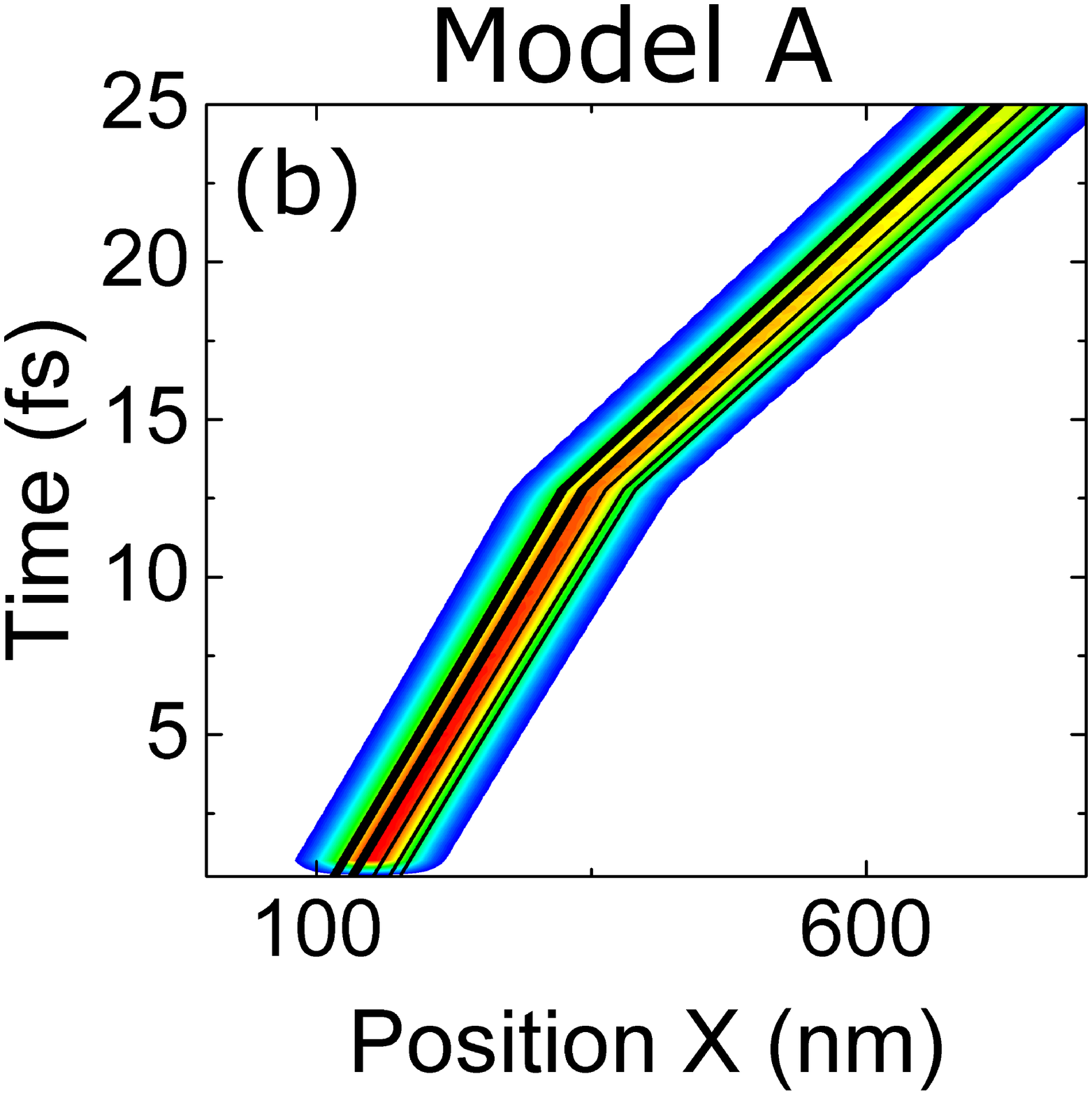}}
	\end{minipage}%
	\vspace{1cm}
	\begin{minipage}{\linewidth}
		\hspace*{-0.3cm}
		\centering
		{ 	\includegraphics[scale=0.28]{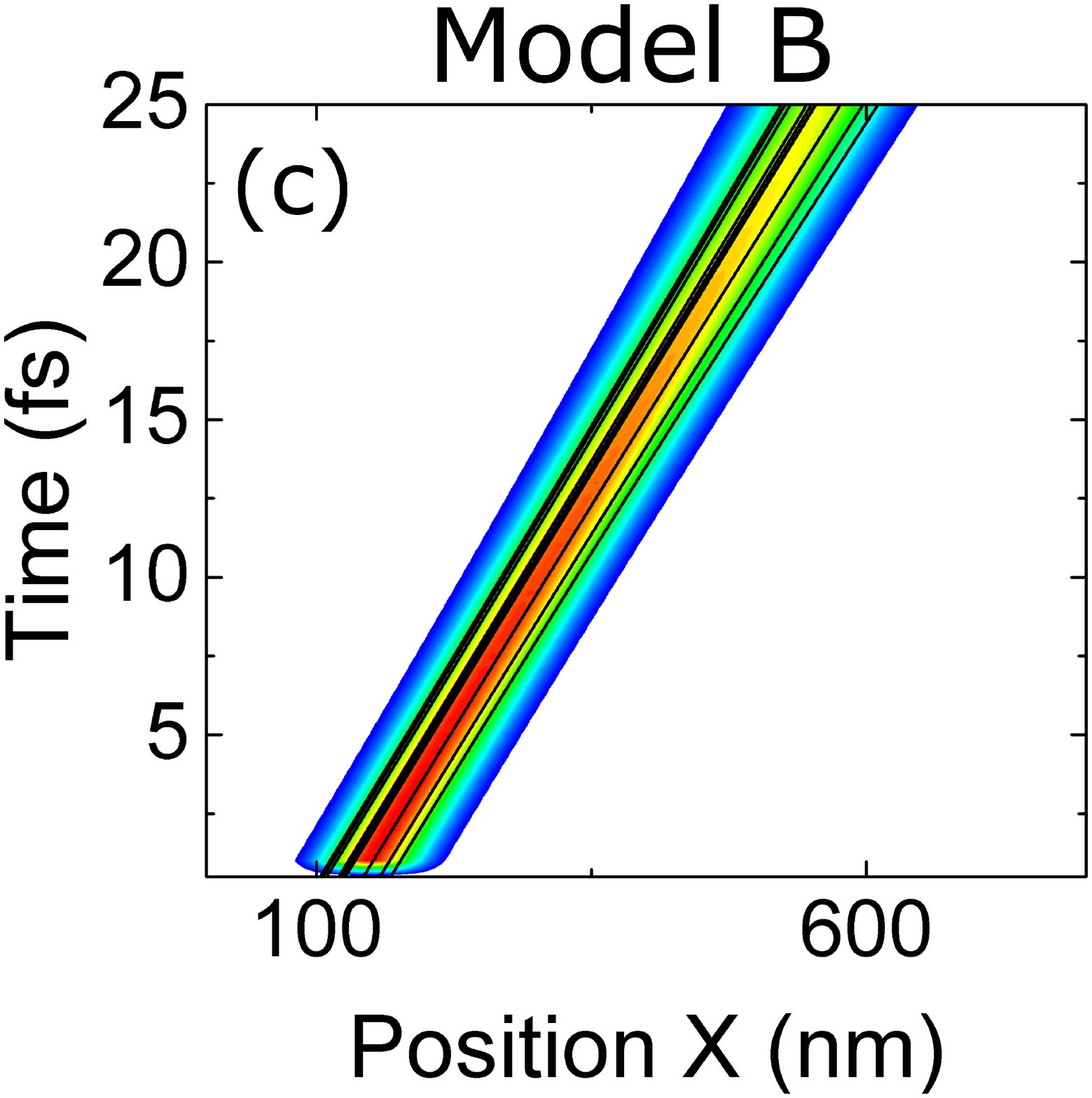}}
		\hspace*{-2.65cm}
		{ 	\includegraphics[scale=0.28]{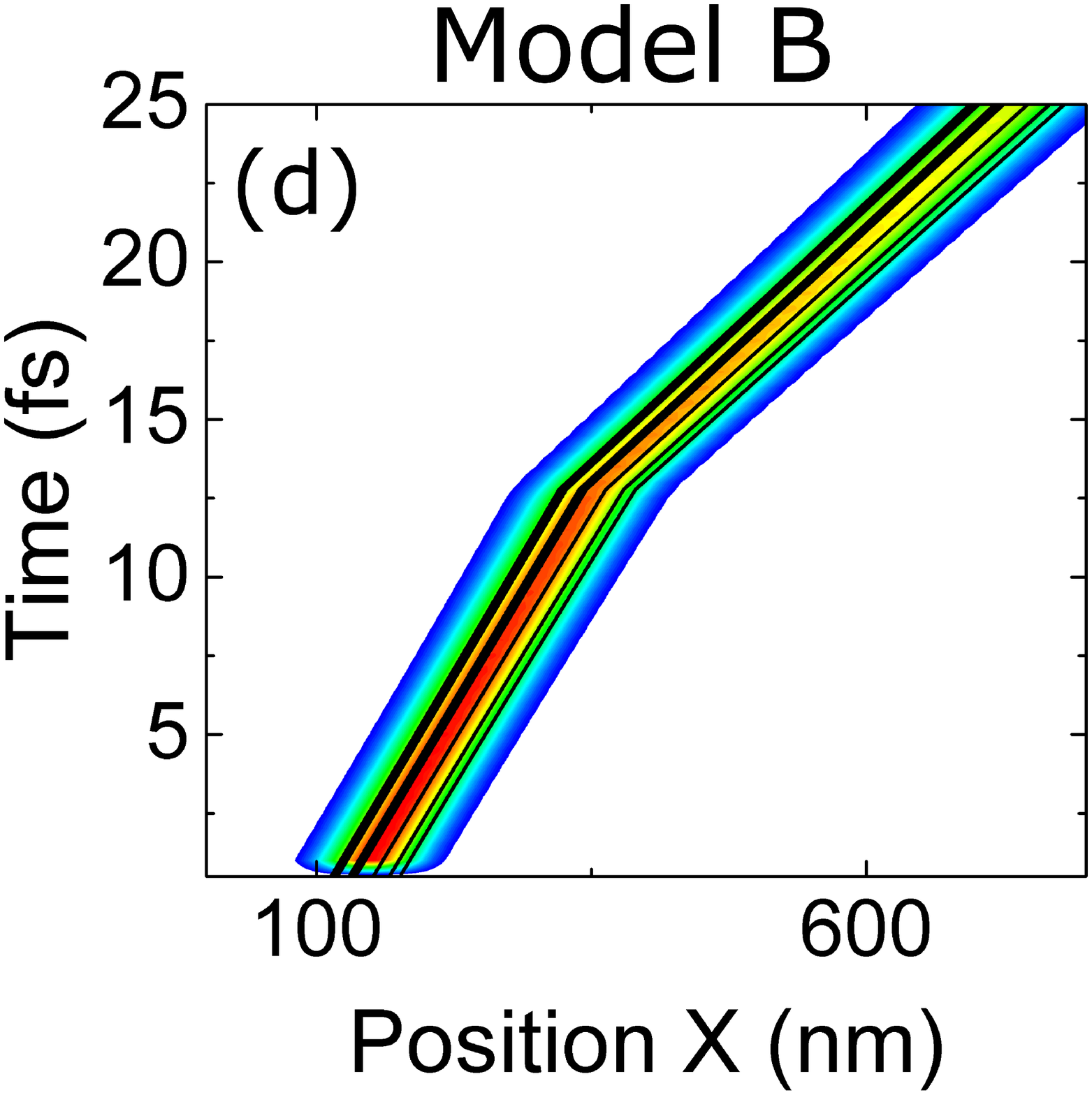}}
	\end{minipage}%
	\caption{The evolution of the BCWF $\psi^j(x,t)$, undergoing photon absorption with $E_\gamma=0.1eV$, shown as function of position and time. The wavefunctions are simulated (a) without collision, (b) with collision using model A, (c) without scattering, (d) scattered using model B. The trajectories $X^j[t]$ guided by the BCWF $\psi^j(x,t)$, where $j=1,\ldots,10$, are some representative experiments and are shown in black. In a flat potential, the results of models A and B are identical.}
	
	\label{comparison_FP_waves_pos}
\end{figure}

\begin{multicols}{2}
In Fig. \ref{comparison_FP_waves_pos} we show the simulation of the electron-photon collision in a flat potential.  The collision is modeled by exchanging the energy $E_\gamma=0.1eV$ in Fig. \ref{comparison_FP_waves_pos}(a) $\to$ (b), and by exchanging the momentum  $p_\gamma=\sqrt{2E_{\gamma}/m^*}$ in Fig. \ref{comparison_FP_waves_pos}(c) $\to$ (d). As expected, in this scenario, both models give identical results.  After the scattering event, the Gaussian wave function evolves with a higher velocity, as indicated in \eqref{extravelo}. We notice that the wave function suffers a continuous evolution during the collision because it is solution of the Schr\"odinger-like equation \eqref{schomomentum}.  Analogous results (not shown) are obtained for emission. The main conclusion of this subsection is that model A and model B are, as expected, numerically equivalent in the case of a flat potential.

\subsection{Collisions in arbitrary potentials}
\label{s5.2}
\end{multicols}
\begin{figure}[H]	
	\begin{minipage}{\linewidth}
		\hspace*{-0.5cm}
		\centering
		{ 	\includegraphics[scale=0.3]{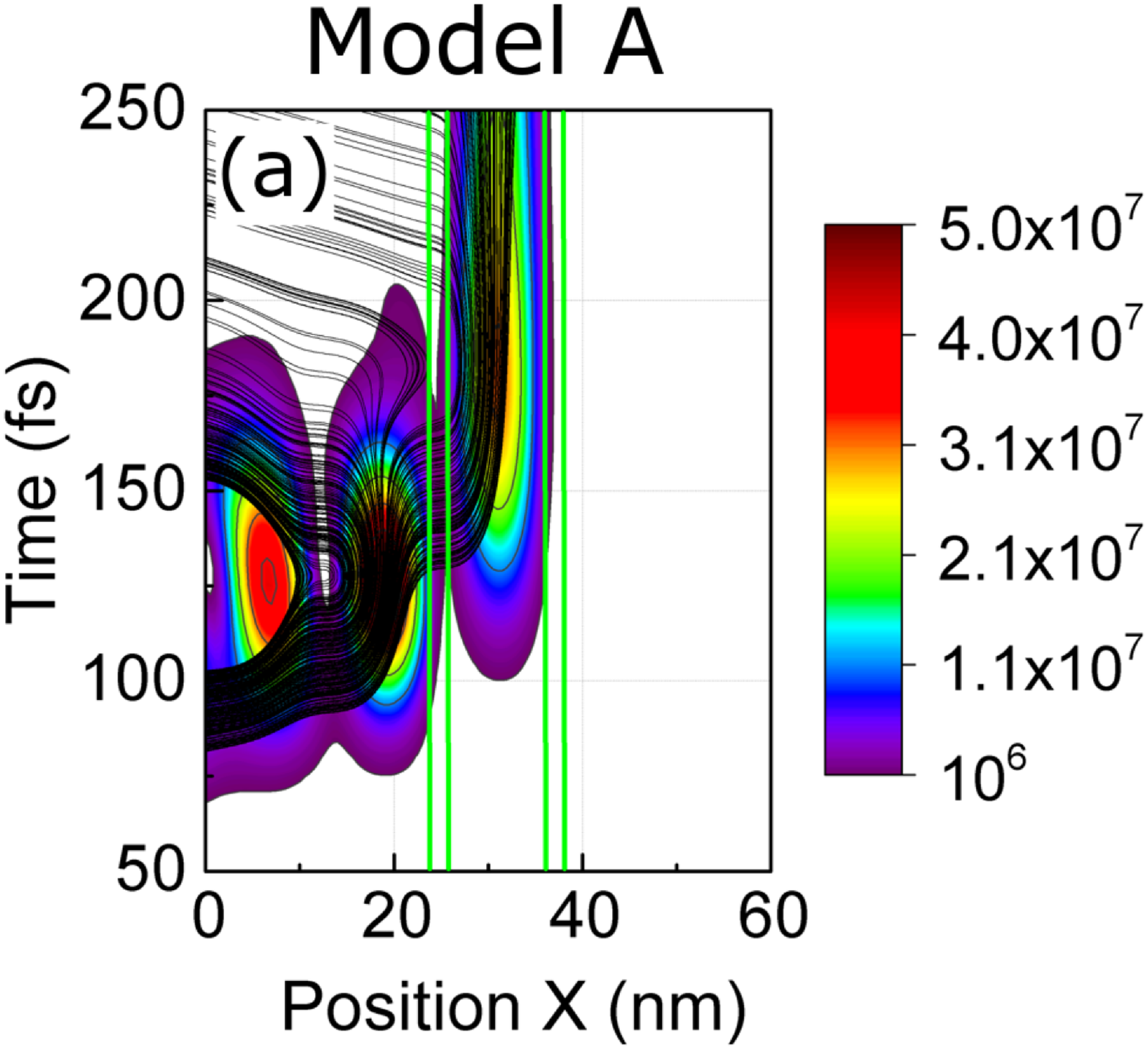}}
		\hspace*{-3.1cm}
		{  \includegraphics[scale=0.3]{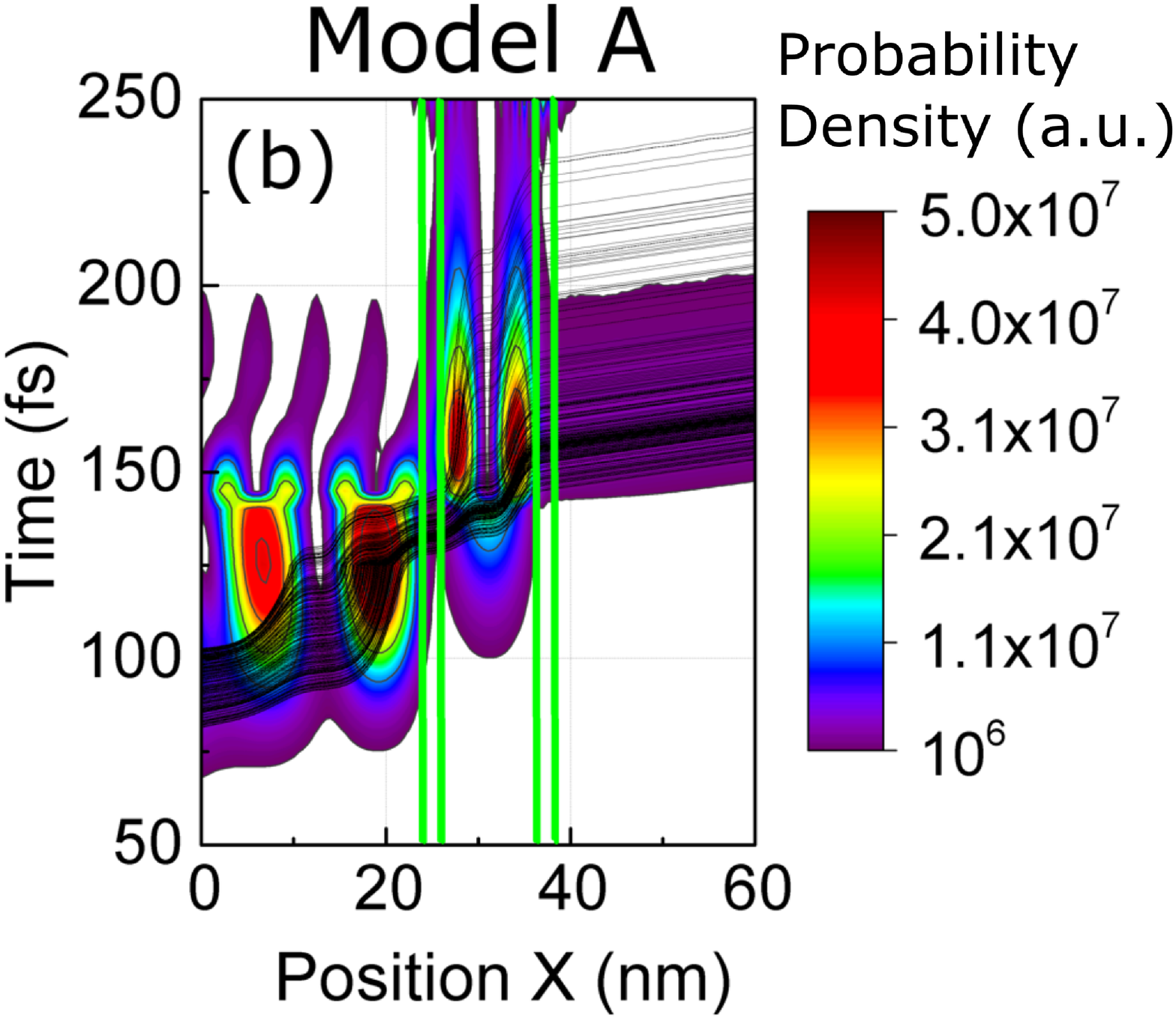}}
		\hspace*{-0.5cm}	
		{ 	\includegraphics[scale=0.3]{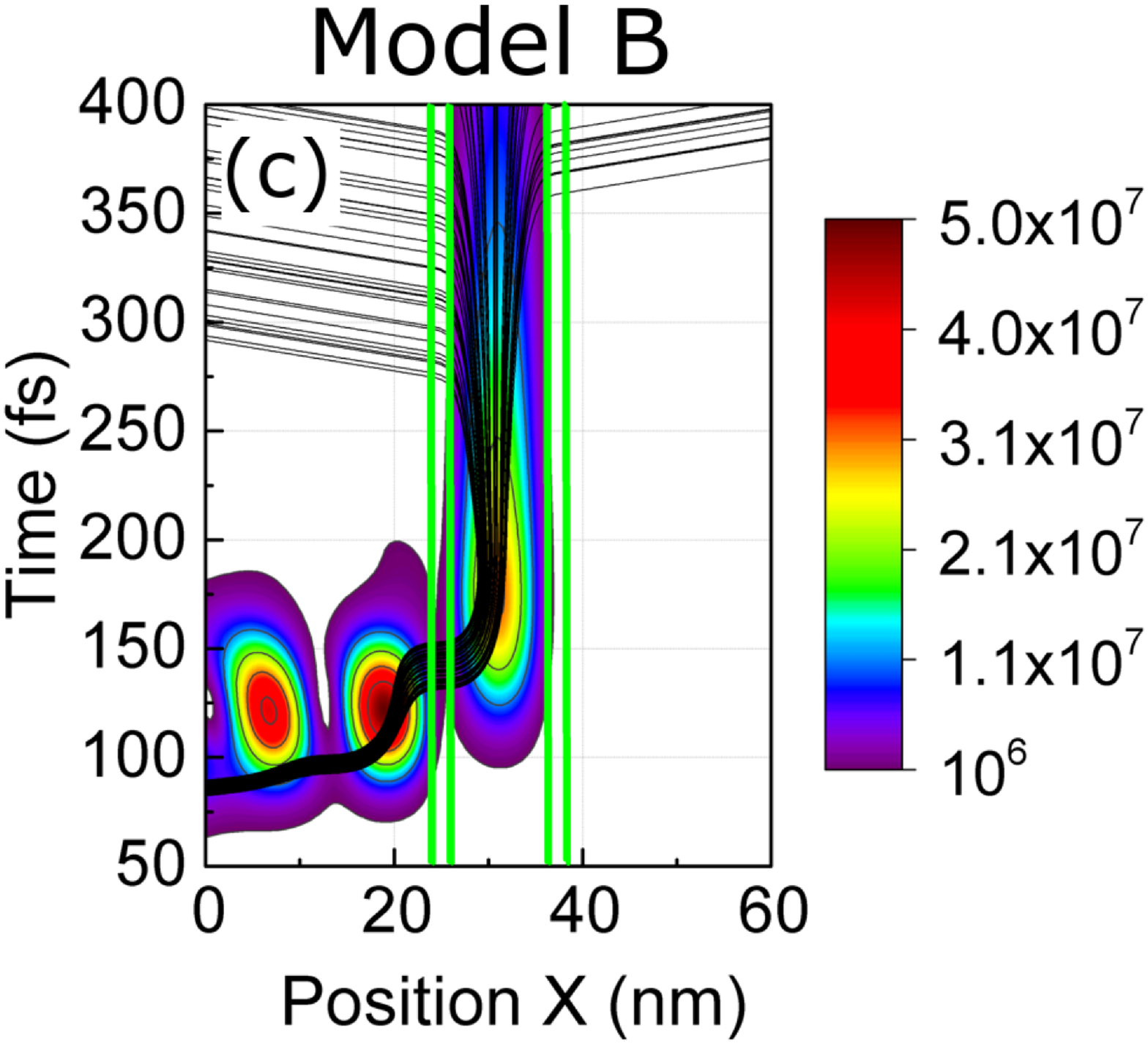}}
		\hspace*{-2.95cm}
		{  \includegraphics[scale=0.3]{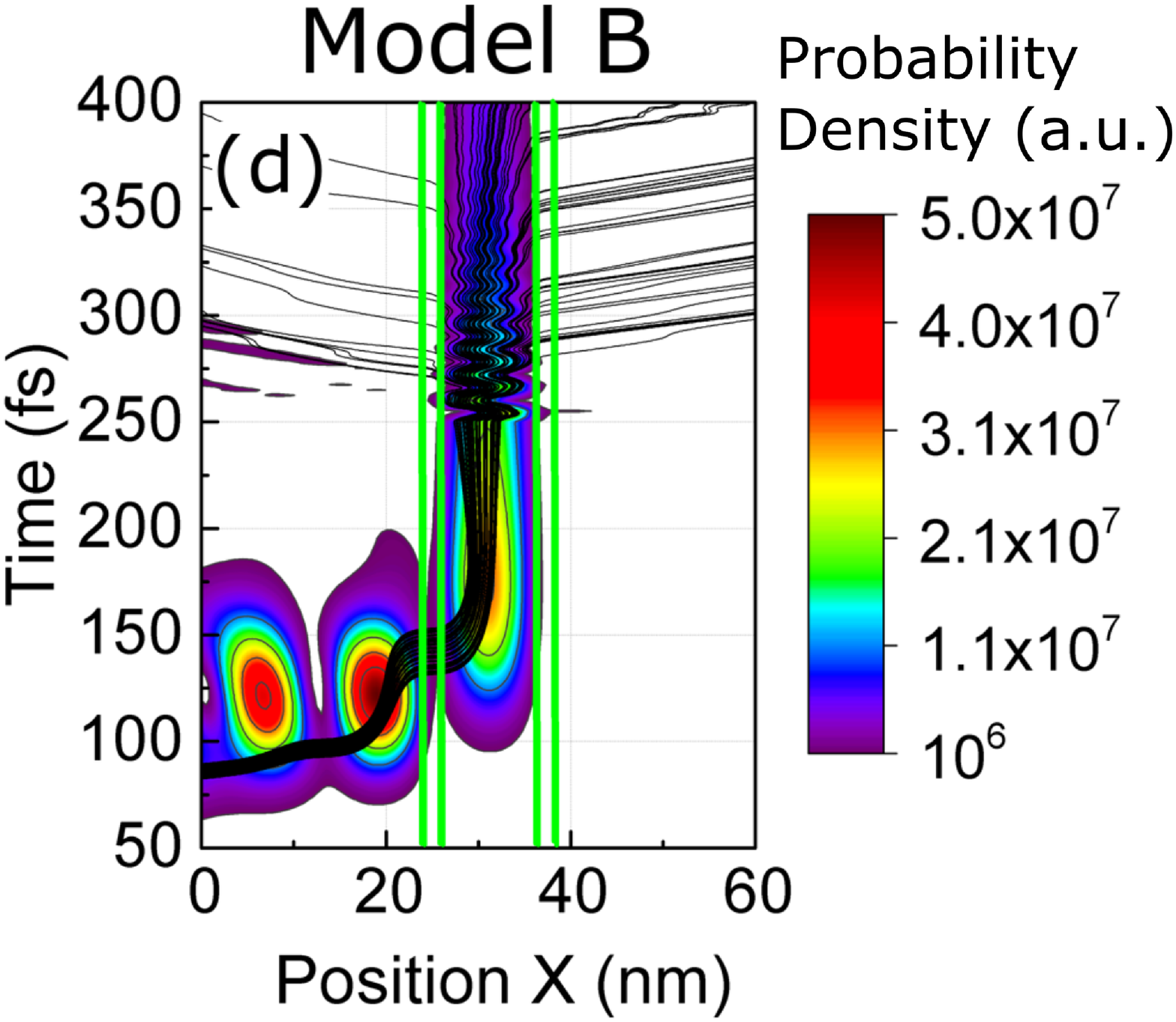}}
	\end{minipage}%
	\caption{Gaussian wavefunctions interacting with a double barrier potential profile with and without scattering with a photon: (a) a wavepacket and some selected trajectories with unitary evolution (without scattering). (b) The same wave packet and the same selected trajectories when scattering with energy $E_\gamma=0.186\,$eV using model A occurs. (c) and (d) are identical to (a) and (b) when model B is used.  In all figures, the Gaussian wave packet is injected from the left at energy $E=E_1=0.058\,$eV. The trajectories $X^j[t]$ guided by the BCWF $\psi^j(x,t)$ are plotted in black. The set of trajectories in plot (a) is different from the one in plot (c), with the goal of selecting those trajectories that most interact in the quantum well in each case. The trajectories in plot (b) are the same as in plot (a) and the trajectories in plot (d) are the same as in plot (c). The energy of the photon is equal to the distance between the two first energy levels, $E_\gamma=E_2-E_1$.}
	\label{wave_comparison_1_to_2}
\end{figure}		
\begin{multicols}{2}

As in Sec. \ref{s5.1}, we study the absorption of a photon by an electron modeled by a final electron (post-selected state) with an energy increase of $\hbar \omega$ ($E_\gamma>0$) with respect to the initial electron energy (pre-selected state). But now we use a double barrier potential $V(x)$ identical to the one mentioned in Sec. \ref{3.1}, with the same two resonant energies $E_1=0.058\,$eV and $E_2=0.23\,$eV.

In Fig.~\ref{wave_comparison_1_to_2} the evolution of $\psi^j(x,t)$ and the trajectories $X^j[t]$ are shown when the electron absorbs a photon while impinging on the potential barrier of the RTD. The position of the barriers is shown by the green vertical lines. The energy of the photon is equal to the difference of the resonant energies in the quantum well, $E_\gamma=E_2-E_1$, and the BCWF is injected with a central energy equal to the first resonant energy $E=E_1$. A transition from $E_1$ to $E_2$ is expected during the collision $\psi_A(x,t_s^-)\to \psi_B(x,t_s^+)$.

In Fig.~\ref{wave_comparison_1_to_2} (a), we plot the time evolution of the electron interacting with the barrier, but without photon collision. In Fig. \ref{wave_comparison_1_to_2} (b), an electron-photon collision is produced at $t_s=150 fs$ using model A. The wavepacket undergoes a shift of the energy probability distribution of the Hamiltonian eigenstates $\phi_E(x)$ towards higher values. As expected, the evolution of $\psi(x,t)$ is a transition from the first eigenstate of the well (with one peak of probability in the middle of the well) to the second one (with two probability peaks). The same trajectories $X^j[t]$ that were first reflected by the barrier in Fig.~\ref{wave_comparison_1_to_2} (a) are now transmitted through the well in Fig. \ref{wave_comparison_1_to_2} (b) because the second resonant level has a wider transmission probability, as shown in Fig. \ref{gig1} (b). The results in Fig. \ref{wave_comparison_1_to_2} (b) have a reasonable agreement with the results in Fig. \ref{evolution_x_exact_model} (a) at times equivalent to the blue and red horizontal lines of Fig. \ref{evolution_x_exact_model} (a). Clearly, we also notice that the simulated result in Fig. \ref{evolution_x_exact_model} (a) belongs to a simulation with the active region as a closed system, where the photon energy does not disappear, an the electron is continuously emitting and absorbing such photon energy, as explained in Sec. \ref{3.1}. On the contrary, Fig. \ref{wave_comparison_1_to_2} (b) corresponds to a simulation of the active region as an open system, where the photon energy appear/disappear at/from the active region only once, as explained in Sec. \ref{3.2}.

\begin{figure}[H]
	\begin{minipage}{\linewidth}
		\centering
		{ 	\includegraphics[width=1\linewidth]{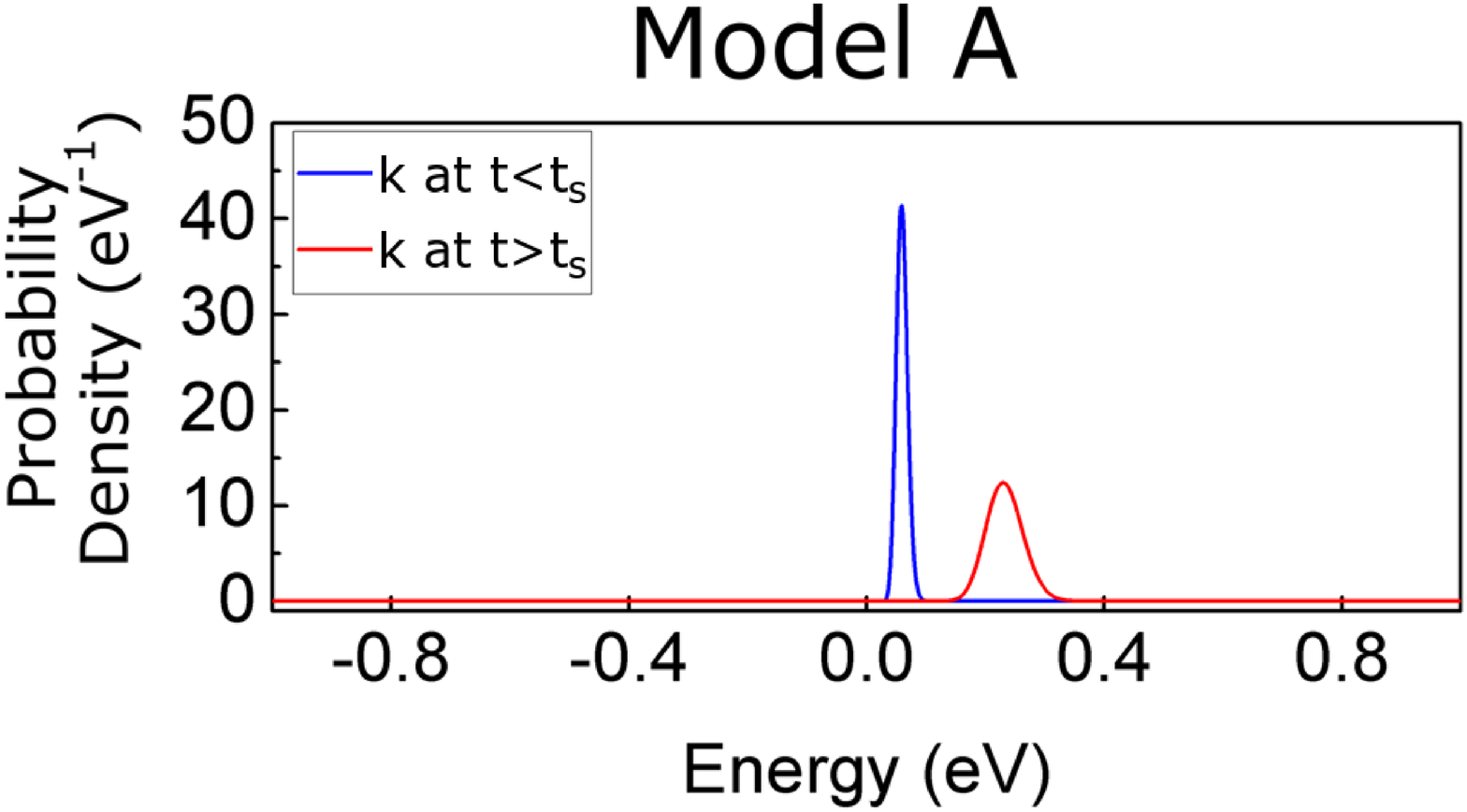}}
		
		{ 	\includegraphics[width=1\linewidth]{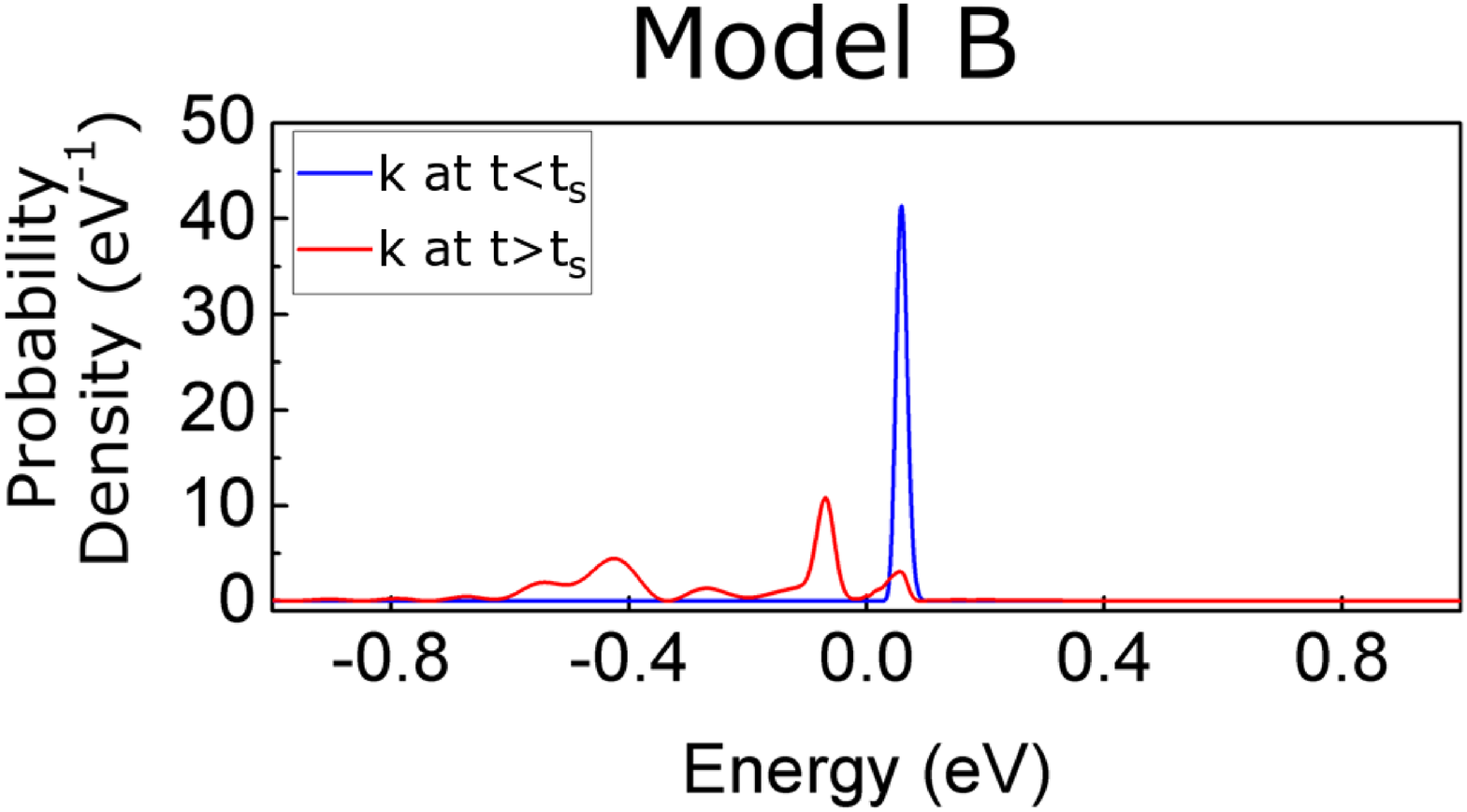}}		
		\caption{(a) Probability distribution of the Hamiltonian eigenstates for Model A (spatial evolution shown in Fig.\ref{wave_comparison_1_to_2}(b)). (b) Probability distribution of the Hamiltonian eigenstates for Model B (spatial evolution shown in Fig.\ref{wave_comparison_1_to_2}(d)). Blue lines represent the probability distribution of the Hamiltonian eigenstates before the scattering at $t<t_s$, while the red lines show it at $t>t_s$.}
		\label{k_old_1_to_2}
	\end{minipage}%
\end{figure}

The same plots are reproduced in Fig. \ref{wave_comparison_1_to_2} (c) and (d) when using model B. Now, an oscillatory behaviour on the BCWF and on the trajectories $X^j[t]$ is shown after time $t_s=250$ fs. Such results can be understood by noticing that model B produces an increase of the velocity in Bohmian trajectories, but such faster Bohmian trajectories are not the \emph{natural} behavior of the trajectories in the well when associated to just one eigenstate (they are expected to remain inside the well for a large time with a velocity close to zero). But since the eigenstates of the quantum well form a complete basis inside the well, the mentioned oscillatory BCWF can be a solution of the Schr\"odinger equation there at the price of using many more eigenstates (with higher energies) to describe the new accelerated wave packet. Thus, the combination of several eigenstates in the well produces the oscillatory behaviour that we see in  Fig.~\ref{wave_comparison_1_to_2}(d).

To better understand that model A provides a \emph{natural} transition, while model B provides an \emph{unnatural} one, we show in Fig.~\ref{k_old_1_to_2} the probability of the energy states $|c(E,t)|^2$ given by Eq.~\eqref{localsuper} at $t=0$ and $t=t_s^+$. The positive and negative energies just indicate scattering states injected from the left (positive) and injected from the right (negative). The blue line is the probability distribution of the energy eigenstates at the initial time $c(E,0)$, while the red line is the same distribution but after the scattering $c(E,t_s^+)$.  In Fig.~\ref{k_old_1_to_2} (a) for model A, we observe a \emph{natural} shift in the central energy given by $\langle E (t_s^+)\rangle=\langle E (t_s^-)\rangle+E_\gamma$, as expected. A definite argument in favor of model A (and against model B) is that the results in Fig.~\ref{k_old_1_to_2} (a) have an almost perfect agreement with the result in Fig. \ref{comparison_FP_neg}(b) that where computed without approximation: the same transition happens from the first to the second energy eigenvalues of the quantum well.
On the contrary, in Fig.~\ref{k_old_1_to_2} (b) for model B, a large amount of Hamiltonian eigenstates with negative energies are \emph{created} after the scattering process. As explained, these additional energy components are the reason why we observe an oscillatory behaviour inside the well in Fig.~\ref{wave_comparison_1_to_2} (d). Model B is nonphysical because it does not satisfy the requirement of conservation of energy in the electron and photon collision. Since we are dealing with a wave packet (with some uncertainty on its energy), some deviation on the requirement of conservation of energy in each experiment is reasonable, but not the deviations plotted in Fig.~\ref{k_old_1_to_2} (b) where energies as high as 1eV are involved.

In conclusion, model B can only describe electron collisions when an approximation of flat potential is reasonable to describe the dynamic of the unperturbed electron. We get exactly the same conclusions when evaluating the emission process (not plotted) instead of the absorption process. In the authors' opinion, this conclusion about model B has dramatic consequences for  quantum transport formalisms that introduce scattering in the position-momentum space, like the Wigner distribution function.

\section{Conclusions}
\label{s6}

Quantum transport formalisms require the modeling of the perturbation induced by the non-simulated degrees of freedom (like photons or phonons) 
on degrees of freedom of the simulated active region (the electrons). 
Among a number of different algorithms that allow to include scattering events, here we explore the possibility of implementing such scattering events as transitions between single-particle pure-states.
We have shown that the Bohmian theory, through the use of BCWF,  allows a rigorous implementation of transitions between pre- and post-selected single-particle pure states in the active device that is valid for both Markovian and non-Markovian conditions.
Furthermore, we have shown that the practical implementation of such transitions requires one to model scattering events as a shift of central energies of BCWFs instead of a shift of central momenta. This last result seems to indicate dramatic consequences for quantum transport formalisms that introduce collisions through changes in momentum, e.g., the Wigner function approach, when dealing with non-flat potential profiles where energy and momentum are non-commuting operators.
The paper is part of a global and long-term research project developing the so-called BITLLES simulator\cite{BITLLES1}. We argue that the amount of information that this simulator framework can provide (from steady-state DC till transient and AC including the fluctuations of the current) in the quantum regime is comparable to the predicting capabilities of the traditional Monte Carlo solution of the Boltzmann transport equation in the semi-classical regime.


\paragraph{Contributions}
``Conceptualization, M.V., X.O., X.C., C.D. and G.A.; methodology, M.V., X.O., X.C., C.D. and G.A.; software, M.V., X.O.; validation, M.V., X.O. X.C., C.D. and G.A.; investigation, M.V., X.O., X.C., C.D. and G.A.; writing---original draft preparation, M.V., X.O., X.C., C.D. and G.A.; writing--review and editing, M.V., X.O., X.C., C.D. and G.A.; visualization, M.V. and X.O.; supervision, X.O., X.C., G.A. and C.D.; project administration, X.O.; funding acquisition, X.O., G.A. and X.C. All authors have read and agreed to the published version of the manuscript.''

\paragraph{Funding}
This research was funded by Spain's Ministerio de Ciencia, Innovaci\'on y Universidades under Grant No. RTI2018-097876-B-C21 (MCIU/AEI/FEDER, UE), the "Generalitat de Catalunya" and FEDER for the project 001-P-001644 (QUANTUMCAT), the European Union's Horizon 2020 research and innovation programme under Grant No. 881603 GrapheneCore3 and under the Marie Sk\l{}odowska-Curie Grant No. 765426 TeraApps.

\paragraph{Conflicts of interest}
The authors declare no conflict of interest. The funders had no role in the design of the study; in the collection, analyses, or interpretation of data; in the writing of the manuscript, or in the decision to publish the results.

\paragraph{Abbreviations}
The following abbreviations are used in this manuscript:\\
	
	\noindent
	\begin{tabular}{@{}ll}
		BCWF & Bohmian Conditional Wave Function\\
		RTD & Resonant Tunnelling Diode
		
\end{tabular}

\end{multicols}
\end{document}